\documentclass[english,british]{article}
\usepackage[utf8]{inputenc}
\usepackage{geometry}
\usepackage{xcolor}
\usepackage{array}
\usepackage{units}
\usepackage{amsthm}
\usepackage{amsmath}
\usepackage{amssymb}
\usepackage{graphicx}
\usepackage{setspace}
\PassOptionsToPackage{normalem}{ulem}
\usepackage{ulem}
\usepackage{subscript}
\usepackage{nameref}

\makeatletter

\providecommand{\tabularnewline}{\\}
\providecolor{lyxadded}{rgb}{0,0,1}
\providecolor{lyxdeleted}{rgb}{1,0,0}

\usepackage{textcomp,marvosym}

\usepackage{fixltx2e}

\usepackage{cite}

\usepackage{nameref,hyperref}

\usepackage{microtype}
\DisableLigatures[f]{encoding = *, family = * }

\usepackage{rotating}

\usepackage[aboveskip=1pt,labelfont=bf,labelsep=period,justification=raggedright,singlelinecheck=off]{caption}

\renewcommand{\@biblabel}[1]{\quad#1.}

\date{}

\@ifundefined{showcaptionsetup}{}{%
 \PassOptionsToPackage{caption=false}{subfig}}
\usepackage{subfig}
\makeatother

\usepackage{babel}
\begin{document}
\vspace*{0.35in}
\begin{flushleft}
{\Large
\textbf\newline{Identification of conserved moieties in metabolic networks by graph theoretical analysis of atom transition networks }
}
\newline
\\
Hulda S. Haraldsdóttir\textsuperscript{1},
Ronan M. T. Fleming\textsuperscript{1,*}
\\
\bigskip
\bf{1} Luxembourg Centre for Systems Biomedicine, University of Luxembourg, Esch-sur-Alzette, Luxembourg
\\
\bigskip

* ronan.mt.fleming@gmail.com

\end{flushleft}

\section*{Abstract}

Conserved moieties are groups of atoms that remain intact in all reactions
of a metabolic network. Identification of conserved moieties gives
insight into the structure and function of metabolic networks and
facilitates metabolic modelling. All moiety conservation relations
can be represented as nonnegative integer vectors in the left null
space of the stoichiometric matrix corresponding to a biochemical
network. Algorithms exist to compute such vectors based only on reaction
stoichiometry but their computational complexity has limited their
application to relatively small metabolic networks. Moreover, the
vectors returned by existing algorithms do not, in general, represent
conservation of a specific moiety with a defined atomic structure.
Here, we show that identification of conserved moieties requires data
on reaction atom mappings in addition to stoichiometry. We present
a novel method to identify conserved moieties in metabolic networks
by graph theoretical analysis of their underlying atom transition
networks. Our method returns the exact group of atoms belonging to
each conserved moiety as well as the corresponding vector in the left
null space of the stoichiometric matrix. It can be implemented as
a pipeline of polynomial time algorithms. Our implementation completes
in under five minutes on a metabolic network with more than 4,000
mass balanced reactions. The scalability of the method enables extension
of existing applications for moiety conservation relations to genome-scale
metabolic networks. We also give examples of new applications made
possible by elucidating the atomic structure of conserved moieties.

\section*{Author summary}

Conserved moieties are transferred between metabolites in internal
reactions of a metabolic network but are not synthesised, degraded
or exchanged with the environment. The total amount of a conserved
moiety in the metabolic network is therefore constant over time. Metabolites
that share a conserved moiety have interdependent concentrations because
their total amount is constant. Identification of conserved moieties
results in a concise description of all concentration dependencies
in a metabolic network. The problem of identifying conserved moieties
has previously been formulated in terms of the stoichiometry of metabolic
reactions. Methods based on this formulation are computationally intractable
for large networks. We show that reaction stoichiometry alone gives
insufficient information to identify conserved moieties. By first
incorporating additional data on the fate of atoms in metabolic reactions,
we developed and implemented a computationally tractable algorithm
to identify conserved moieties and their atomic structure.

\section{\label{sec:Introduction}Introduction}

Conserved moieties give rise to pools of metabolites with constant
total concentration and dependent individual concentrations. These
constant metabolite pools often consist of highly connected cofactors
that are distributed throughout a metabolic network. Representative
examples from energy metabolism include the AMP and NAD moieties \cite{atkinson_cellular_1977,reich_energy_1981}.
Changes in concentration ratios within these cofactor pools affect
thermodynamic and mass action kinetic driving forces for all reactions
they participate in. Moiety conservation therefore imposes a purely
physicochemical form of regulation on metabolism that is mediated
through changes in concentration ratios within constant metabolite
pools. Reich and Sel'kov likened conserved moieties to turning wheels
that are ``geared into a clockwork'' \cite{reich_energy_1981}.
They described the thermodynamic state of energy metabolism as ``open
flow through a system closed by moiety conservation''. Identification
of conserved moieties in metabolic networks has helped elucidate complex
metabolic phenomena including synchronisation of glycolytic oscillations
in yeast cell populations \cite{bier_how_2000} and the function of
glycosomes in the African sleeping sickness parasite \textit{Trypanosoma
brucei} \cite{bakker_compartmentation_2000}. It has also been shown
to be relevant for drug development \cite{bakker_compartmentation_2000,cornish-bowden_role_2002}.

Identification of conserved moieties has been of interest to the metabolic
modelling community for several decades \cite{sauro_conservation_2004,vallabhajosyula_conservation_2006}.
It is particularly important for dynamic modelling \cite{horn_general_1972}
and metabolic control analysis \cite{hofmeyr_metabolic_1986} where
metabolite concentrations are explicitly modelled. Moiety conservation
relations provide a sparse, physically meaningful description of concentration
dependencies in a metabolic network. They can be used to eliminate
redundant metabolite concentrations as the latter can be derived from
the set of independently varying metabolite concentrations. Doing
so facilitates simulation of metabolic networks and is in fact required
for many computational modelling methods \cite{sauro_conservation_2004,vallabhajosyula_conservation_2006}.

\begin{figure}[h]
\begin{centering}
\includegraphics{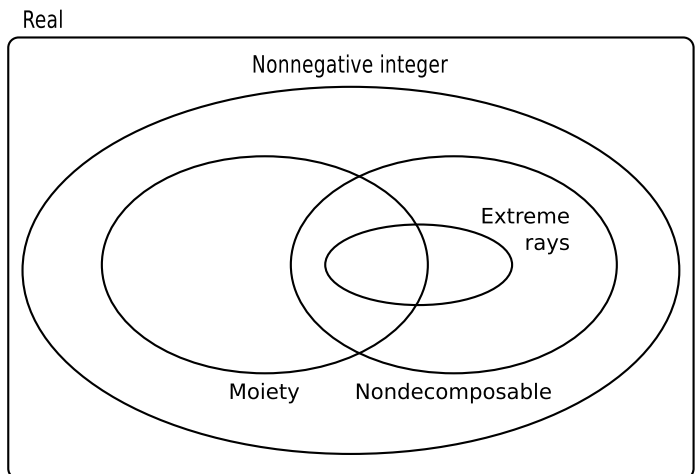}
\par\end{centering}

\protect\caption{\label{fig:Sets}\textbf{Sets of conservation vectors for metabolic
networks.} The set of real-valued conservation vectors consists of
all vectors in the left null space of a stoichiometric matrix. Real-valued
basis vectors can be computed using efficient linear algebra algorithms
but are difficult to interpret as they generally contain negative
and noninteger elements. Nonnegative integer vectors are easier to
interpret but more difficult to compute. Existing algorithms have
exponential worst case time complexity. Algorithms exist to compute
extreme rays, the set of all nondecomposable nonnegative integer vectors,
and a maximal set of linearly independent nonnegative integer vectors.
These vector sets intersect with the set of moiety vectors but are
not equivalent to it. Moiety vectors represent conservation of an
identifiable group of atoms in network metabolites. They are a property
of the specific set of metabolites and reactions that constitute a
metabolic network whereas other conservation vectors are a property
of the network's stoichiometric matrix. The method presented here
computes moiety vectors in polynomial time.}
\end{figure}

Mathematically, moiety conservation gives rise to a stoichiometric
matrix with linearly dependent rows. The left null space of the stoichiometric
matrix therefore has nonzero dimension (see Section \ref{sub:moieties}). Vectors in the left null space,
hereafter referred to as conservation vectors, can be divided into
several interrelated sets based on their numerical properties and
biochemical meaning (Fig. \ref{fig:Sets}). \textit{Moiety vectors}
constitute a subset of conservation vectors with a distinct biochemical
interpretation. Each moiety vector represents conservation of a particular
metabolic moiety. Elements of a moiety vector correspond to the number
of instances of a conserved moiety in metabolites of a metabolic network.
As moieties are discrete quantities, moiety vectors are necessarily
nonnegative integer vectors.

Methods exist to compute conservation vectors based only on the stoichiometric
matrix of a metabolic network. These methods compute different types
of bases for the left null space of the stoichiometric matrix (see
\nameref{sub:S1_Appendix} for mathematical definitions). Each method
draws basis vectors from a particular set of conservation vectors
(Fig. \ref{fig:Sets}). There is a tradeoff between the computational
complexity of these methods and the biochemical interpretability of
the basis vectors they return. At the low end of the computational
complexity spectrum are linear algebraic methods such as singular
value decomposition. Other methods, such as Householder QR factorisation
\cite{vallabhajosyula_conservation_2006} or sparse LU factorisation
\cite{gill_maintaining_1987} are more efficient for large stoichiometric
matrices. These methods construct a \textit{linear basis} for the
left null space from real-valued conservation vectors. Though readily
computed, these vectors are also the most difficult to interpret as
they generally contain negative and noninteger elements.

Schuster and Höfer \cite{schuster_determining_1991} introduced the
use of vertex enumeration algorithms to compute the \textit{extreme
rays} of the positive orthant of the left null space. They referred
to these extreme rays as ``extreme semipositive conservation relations''.
Famili and Palsson \cite{famili_convex_2003} later referred to them
as ``metabolic pools'' and the set of all extreme rays as ``a \textit{convex
basis} for the left null space''. Like moiety vectors, extreme rays
are nonnegative integer vectors. They are therefore readily interpreted
in terms of constant metabolite pools. However, extreme rays can currently
only be computed for relatively small metabolic networks due to the
computational complexity of vertex enumeration algorithms \cite{avis_pivoting_1992}.
Moreover, the set of extreme rays is not identical to the set of moiety
vectors (Fig. \ref{fig:Sets}). Schuster and Hilgetag \cite{schuster_what_1995}
presented examples of extreme rays that did not represent moiety conservation
relations, as well as moiety vectors that were not extreme rays.

Moiety vectors are a property of a metabolic network while extreme
rays are a property of its stoichiometric matrix. Multiple metabolic
networks could in theory have the same stoichiometric matrix, despite
consisting of different sets of metabolites and reactions. These networks
would all have the same set of extreme rays, but could have different
sets of moiety vectors. Schuster and Hilgetag \cite{schuster_what_1995}
published an extension to the vertex enumeration algorithm in \cite{schuster_determining_1991}
to compute the set of all \textit{nondecomposable nonnegative integer
vectors} in the left null space of a stoichiometric matrix. This set
is guaranteed to contain all nondecomposable moiety vectors for a
particular metabolic network as subset (Fig. \ref{fig:Sets}). However,
it is impossible to identify the subset of moiety vectors without
information about the atomic structure of metabolites.

Alternatives to vertex enumeration have been proposed to speed up
computation of biochemically meaningful conservation vectors, e.g.,
\cite{nikolaev_elucidation_2005,soliman_invariants_2012,de_martino_identifying_2014}.
Most recently, De Martino et al. \cite{de_martino_identifying_2014}
published a novel method to compute a \textit{nonnegative integer
basis} for the left null space of a stoichiometric matrix. This method
\cite{de_martino_identifying_2014} relies on stochastic algorithms,
without guaranteed convergence, but that were empirically shown to
perform well even on large networks. Like extreme rays, the nonnegative
integer vectors computed with this method are not necessarily moiety
vectors (Fig. \ref{fig:Sets}). In general, methods to analyse stoichiometric
matrices are not suited to specifically compute moiety vectors. Computation
of moiety vectors requires information about the atomic composition
of metabolites. To our knowledge, only one method has previously been
published to specifically compute moiety vectors for metabolic networks
\cite{park_jr._complete_1986}. This method was based on nonnegative
integer factorisation of the elemental matrix; a numerical representation
of metabolite formulas. Nonnegative integer factorisation of a matrix
is at least NP-hard \cite{vavasis_complexity_2007} and no polynomial
time algorithm is known to exist for this problem. Moreover, only
the chemical formula but not the atomic identities of the conserved
moieties can be derived from this approach. Identifying the atoms
that belong to each moiety requires additional information about the
fate of atoms in metabolic reactions. This information is not contained
in a stoichiometric matrix.

Here, we propose a novel method to identify conserved moieties in
metabolic networks. Our method is based on the premise that atoms
within the same conserved moiety follow identical paths through a
metabolic network. Given data on which substrate atoms map to which
product atoms in each metabolic reaction, the paths of individual
atoms through a metabolic network can be encoded in an \textit{atom
transition network}. Until recently, the necessary data were difficult
to obtain but relatively efficient algorithms have now become available
to predict atom mappings in metabolic reactions \cite{first_stereochemically_2012,latendresse_accurate_2012,kumar_clca:_2014}.
These algorithms have made it possible to construct atom transition
networks for large metabolic networks. Unlike metabolic networks,
atom transition networks are amenable to analysis with efficient graph
theory algorithms. Here, we take advantage of this fact to identify
conserved moieties in metabolic networks in polynomial time. Furthermore,
starting from atom transition networks allows us to associate each
conserved moiety with a specific group of atoms in a subset of metabolites
in a metabolic network.

This work combines elements of biochemistry, linear algebra and graph
theory. We have made an effort to accommodate readers from all fields.
The main text consists of informal descriptions of our methods and
results, accompanied by illustrative examples and a limited number
of mathematical equations. Formal definitions of italicised terms
are given in supporting file \nameref{sub:S1_Appendix}. We precede
our results with a section on the theoretical framework for this work,
where we introduce key concepts and notation used in the remainder
of the text.

\section{\label{sec:Framework}Theoretical framework}

\subsection{\label{sub:Met}Metabolic networks}

A metabolic network consists of a set of metabolites that interconvert
via a set of metabolic reactions. Metabolic networks in living beings
are open systems that exchange mass and energy with their environment.
For modelling purposes, the boundary between system and environment
can be defined by introducing a set of metabolite sources and sinks
collectively known as exchange reactions. Unlike internal reactions,
exchange reactions are artificial constructs that do not conserve
mass or charge. The topology of a metabolic network can be represented
in several ways. Here, we use metabolic maps and stoichiometric matrices.
A metabolic map for a small example metabolic network is shown in
Fig. \ref{fig:ex_met}. This example will be used throughout this
section to demonstrate key concepts relevant to this work.

\begin{center}
\begin{figure}[h]
\begin{centering}
\includegraphics{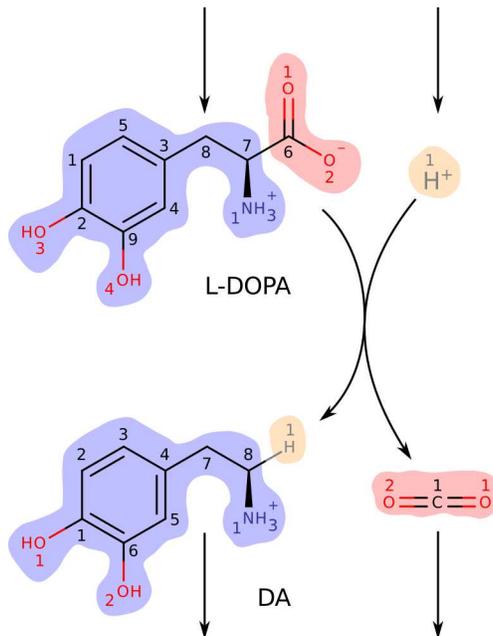}
\par\end{centering}

\protect\caption{\label{fig:ex_met}\textbf{A metabolic map for an example metabolic
network.} The network consists of one internal reaction and four exchange
reactions. The internal reaction is the DOPA decarboxylase reaction
(KEGG Reaction ID: R02080) that produces dopamine (DA, KEGG Compound
ID: C03758) and CO\protect\textsubscript{2} from levodopa (L-DOPA,
C00355) and H\protect\textsuperscript{+}. The open network includes
source reactions for the two substrates and sink reactions for the
two products. Arrowheads indicate reaction directionality. Metabolite
structures were rendered from molfiles (Accelrys, San Diego, CA) with
MarvinView (ChemAxon, Budapest, Hungary). Atoms are numbered according
to their order in each metabolite's molfile. Atoms of different elements
are numbered separately, in colours matching their elemental symbol.
The internal reaction conserves three metabolic moieties. Atoms belonging
to the same moiety have identically coloured backgrounds. Levodopa
and dopamine each contain one instance of a dopamine moiety (blue
background). Implicit hydrogen atoms on both metabolites are also
part of this moiety. Levodopa and CO\protect\textsubscript{2} each
contain one instance of a CO\protect\textsubscript{2} moiety (red
background). Finally, the hydrogen ion and dopamine each contain one
instance of a hydrogen moiety (orange background).}
\end{figure}

\par\end{center}

A stoichiometric matrix for an open metabolic network with $m$ metabolites
and $n$ reactions is denoted by $S\in\mathbb{R}^{m\times n}$. Each
row of $S$ represents a metabolite and each column a reaction such
that element $S_{ij}$ is the stoichiometric coefficient of metabolite
$i$ in reaction $j$. Coefficients are negative for substrates and
positive for products. Substrates and products in reversible reactions
are defined by designating one direction as forward. The stoichiometric
matrix can be written as 

\begin{equation}
S=\left[N,B\right],
\end{equation}
where $N\in\mathbb{Z}^{m\times u}$ consists of columns representing
internal (mass balanced) reactions and $B\in\mathbb{R}^{m\times\left(n-u\right)}$
consists of columns representing exchange reactions (mass imbalanced).
Note that $N$ represents a metabolic network that is closed to the
environment. In what follows we will refer to $N$ as the internal
stoichiometric matrix, $B$ as the exchange stoichiometric matrix,
and $S$ as the total stoichiometric matrix. The total stoichiometric
matrix for the example metabolic network in Fig. \ref{fig:ex_met}
is given in Table \ref{tab:ex_S}.

\begin{center}
\begin{table}[!th]
\protect\caption{\label{tab:ex_S}\textbf{The total stoichiometric matrix $S=\left[N,B\right]$
for the example metabolic network.}}

\begin{centering}
\begin{tabular}{l|lc|ccccc|}
\multicolumn{1}{l}{} &  & N & \multicolumn{4}{c}{B} & \multicolumn{1}{c}{}\tabularnewline
\cline{2-2} \cline{8-8} 
L-DOPA &  & -1 & 1 & 0 & 0 & 0 & \tabularnewline
H\textsuperscript{+} &  & -1 & 0 & 1 & 0 & 0 & \tabularnewline
DA &  & 1 & 0 & 0 & -1 & 0 & \tabularnewline
CO\textsubscript{2} &  & 1 & 0 & 0 & 0 & -1 & \tabularnewline
\cline{2-2} \cline{8-8} 
\end{tabular}\bigskip{}

\par\end{centering}

Rows are labelled with the corresponding metabolite identifier from
Fig. \ref{fig:ex_met}. The internal stoichiometric matrix $N\in\mathbb{Z}^{4\times1}$
is row rank deficient, with ${\rm rank}\left(N\right)=1$. The dimension
of its left null space is therefore $\dim\left({\cal N}\left(N^{T}\right)\right)=4-1=3$.
The total stoichiometric matrix $S\in\mathbb{Z}^{4\times5}$ is full
row rank. Its left null space is therefore zero dimensional.
\end{table}

\par\end{center}

Stoichiometric matrices are \textit{incidence matrices} for generalised
\textit{graphs} known as \textit{hypergraphs} \cite{klamt_hypergraphs_2009}.
Hypergraphs contain hyperedges that can connect more than two nodes.
The metabolic map in Figure \ref{fig:ex_met} is a planar visualisation
of a hypergraph with one hyperedge, connecting four metabolites. A
graph edge that only connects two nodes is a special instance of a
hyperedge. Apart from the occasional isomerisation reaction, metabolic
reactions involve more than two metabolites. As a result, they cannot
be represented as graph edges without loss of information. Metabolic
networks are therefore represented as hypergraphs where nodes represent
metabolites and hyperedges represent reactions. Since reactions have
a designated forward direction, they are \textit{directed hypergraphs}.
Representing metabolic networks as hypergraphs has the advantage of
conserving basic structure and functional relationships. The disadvantage
is that many graph theoretical algorithms are not applicable to hypergraphs
\cite{klamt_hypergraphs_2009}.

\subsection{\label{sub:moieties}Moiety vectors}

An internal stoichiometric matrix $N\in\mathbb{Z}^{m\times u}$ for
a closed metabolic network is always row-rank deficient, i.e., $\textrm{rank}\left(N\right)<m$
\cite{schuster_determining_1991}. The left null space of $N$, denoted
by ${\cal N}\left(N^{T}\right)$, therefore has finite dimension given
by $\dim\left({\cal N}\left(N^{T}\right)\right)=m-\textrm{rank}\left(N\right)$.
The left null space holds all conservation vectors for a stoichiometric
matrix \cite{horn_general_1972}. The number of linearly independent
conservation vectors for a closed metabolic network is $\dim\left({\cal N}\left(N^{T}\right)\right)$.
The total stoichiometric matrix $S$ for an open metabolic network
has a greater rank than the internal stoichiometric matrix $N$ for
the corresponding closed metabolic network (e.g., Table \ref{tab:ex_S}),
i.e., ${\rm rank}\left(N\right)<{\rm rank}\left(S\right)$. Consequently,
$\dim\left({\cal N}\left(S^{T}\right)\right)<\dim\left({\cal N}\left(N^{T}\right)\right)$,
meaning that there are fewer linearly independent conservation vectors
for an open metabolic network than the corresponding closed network.
This is consistent with physical reality, since mass can flow into
and out of open networks but is conserved within closed networks.
All quantities that are conserved in an open metabolic network are
also conserved in the corresponding closed network. That is, if $z$
is a conservation vector for an open metabolic network $S$, such
that $S^{T}z=0$, then $z$ is also a conservation vector for the
corresponding closed network $N$, and $N^{T}z=0$, since $S=\left[N,B\right]$.
The set of conservation relations for an open network is therefore
a subset of all conservation relations for the corresponding closed
network, i.e., ${\cal N}\left(S^{T}\right)\subseteq{\cal N}\left(N^{T}\right)$.
In what follows we will mainly be concerned with the larger set of
conservation relations for a closed metabolic network.

Schuster and Hilgetag \cite{schuster_what_1995} defined a moiety
vector $l_{1}$ as a nonnegative integer vector in the left null space
of a stoichiometric matrix, i.e., 
\begin{equation}
N^{T}l_{1}=0,\label{eq:p1}
\end{equation}
\begin{equation}
l_{1}\in\mathbb{N}_{0}^{m}.\label{eq:p2}
\end{equation}
In addition, they defined $l_{1}$ to be a maximal moiety vector if
it cannot be decomposed into two other vectors $l_{2}$ and $l_{3}$
that satisfy Eq. \ref{eq:p1} and \ref{eq:p2}, i.e., if

\begin{equation}
l_{1}\neq\alpha_{2}l_{2}+\alpha_{3}l_{3},\label{eq:p3}
\end{equation}
where $\alpha_{2},\alpha_{3}\in\mathbb{N}_{+}$. We propose a more
specific definition. The properties above define increasingly small
sets of conservation vectors (Fig. \ref{fig:Sets}). Equation \ref{eq:p1}
defines the set of all conservation vectors. Addition of Eq. \ref{eq:p2}
defines the set of nonnegative integer conservation vectors and addition
of Eq. \ref{eq:p3} defines the set of nonnegative integer conservation
vectors that are nondecomposable. Although this set includes all nondecomposable
moiety vectors as subset it is not equivalent (Fig. \ref{fig:Sets}).
To define the set of moiety vectors we require a fourth property.
We define $l_{1}$ to be a moiety vector if it satisfies Eq. \ref{eq:p1}
and \ref{eq:p2} and represents conservation of a specific metabolic
moiety, i.e., an identifiable group of atoms in network metabolites.
Element $l_{1,i}$ should correspond to the number of instances of
the conserved moiety in metabolite $i$. We define $l_{1}$ to be
a \textit{nondecomposable moiety vector} if it satisfies condition
\ref{eq:p3} and a \textit{composite moiety vector} if it does not.
Nondecomposable moiety vectors for the DOPA decarboxylase reaction
from the example metabolic network in Fig. \ref{fig:ex_met} are given
in Table \ref{tab:MoietyVectors}. For comparison, conservation vectors
computed with existing methods for conservation analysis of metabolic
networks are given in Tables \ref{tab:E}-\ref{tab:ExtremeRays}.
In general, these vectors do not represent moiety conservation.

\begin{table}[!th]
\protect\caption{\textbf{\label{tab:ex_vectors}Different types of conservation vectors
for the DOPA decarboxylase reaction.}}

\begin{centering}
\subfloat[\label{tab:MoietyVectors}]{

\centering{}%
\begin{tabular}{l|lrrrc|}
\multicolumn{1}{l}{} &  & $l_{1}$ & $l_{2}$ & $l_{3}$ & \multicolumn{1}{c}{}\tabularnewline
\cline{2-2} \cline{6-6} 
L-DOPA &  & 1 & 1 & 0 & \tabularnewline
H\textsuperscript{+} &  & 0 & 0 & 1 & \tabularnewline
DA &  & 1 & 0 & 1 & \tabularnewline
CO\textsubscript{2} &  & 0 & 1 & 0 & \tabularnewline
\cline{2-2} \cline{6-6} 
\end{tabular}}\subfloat[\label{tab:E}]{

\centering{}%
\begin{tabular}{|lccccc|}
\multicolumn{1}{l}{} & C & H & O & N & \multicolumn{1}{c}{}\tabularnewline
\cline{1-1} \cline{6-6} 
 & 9 & 11 & 4 & 1 & \tabularnewline
 & 0 & 1 & 0 & 0 & \tabularnewline
 & 8 & 12 & 2 & 1 & \tabularnewline
 & 1 & 0 & 2 & 0 & \tabularnewline
\cline{1-1} \cline{6-6} 
\end{tabular}}\subfloat[\label{tab:SVD}]{

\centering{}%
\begin{tabular}{|lcccc|}
\multicolumn{1}{l}{} & ~~$z_{1}$ & ~~$z_{2}$ & ~~$z_{3}$ & \multicolumn{1}{c}{}\tabularnewline
\cline{1-1} \cline{5-5} 
 & $-\nicefrac{1}{2}$ & ~~$\nicefrac{1}{2}$ & ~~$\nicefrac{1}{2}$ & \tabularnewline
 & ~~$\nicefrac{5}{6}$ & ~~$\nicefrac{1}{6}$ & ~~$\nicefrac{1}{6}$ & \tabularnewline
 & ~~$\nicefrac{1}{6}$ & ~~$\nicefrac{5}{6}$ & $-\nicefrac{1}{6}$ & \tabularnewline
 & ~~$\nicefrac{1}{6}$ & $-\nicefrac{1}{6}$ & ~~$\nicefrac{5}{6}$ & \tabularnewline
\cline{1-1} \cline{5-5} 
\end{tabular}}\subfloat[\label{tab:ExtremeRays}]{

\centering{}%
\begin{tabular}{|lrrrrc|}
\multicolumn{1}{l}{} & $l_{1}$ & $l_{2}$ & $l_{3}$ & $z_{4}$ & \multicolumn{1}{c}{}\tabularnewline
\cline{1-1} \cline{6-6} 
 & 1 & 1 & 0 & 0 & \tabularnewline
 & 0 & 0 & 1 & 1 & \tabularnewline
 & 1 & 0 & 1 & 0 & \tabularnewline
 & 0 & 1 & 0 & 1 & \tabularnewline
\cline{1-1} \cline{6-6} 
\end{tabular}}
\par\end{centering}

Moiety vectors are denoted $l_{k}$. (a) Moiety vectors computed with
the method presented here. Each column represents the conservation
of a particular metabolic moiety. $l_{1}$ represents conservation
of the dopamine moiety (blue background in Fig. \ref{fig:ex_met}),
$l_{2}$ the CO\textsubscript{2} moiety (red background), and $l_{3}$
the hydrogen moiety (orange background). (b) The elemental matrix.
Each column represents conservation of a particular element. Elemental
conservation vectors generally do not span the left null space of
a stoichiometric matrix. (c) Real-valued conservation vectors computed
with singular value decomposition of the internal stoichiometric matrix
$N$ in Table \ref{tab:ex_S}. Real-valued conservation vectors cannot
generally be interpreted in terms of conserved moieties as they contain
negative and noninteger values. (d) Extreme rays of the left null
space $\mathcal{N}\left(N^{T}\right)$. The first three belong to
the intersection between the sets of extreme rays and moiety vectors
in Fig. \ref{fig:Sets}. The fourth belongs to the set difference.
It cannot represent moiety conservation as no atoms are exchanged
between H\textsuperscript{+} and CO\textsubscript{2}. Without information
about atom mappings between metabolites it would be impossible to
determine which extreme rays correspond to conserved moieties. The
full set of all nondecomposable nonnegative integer vectors includes
13 additional vectors (not shown), none of which represent moiety
conservation.
\end{table}

\subsection{\label{sub:ATNs}Atom transition networks}

\begin{figure}[h]
\begin{centering}
\includegraphics{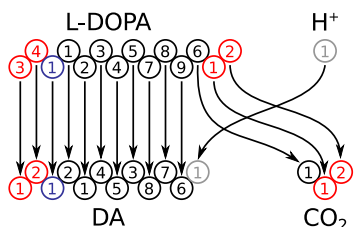}
\par\end{centering}

\protect\caption{\label{fig:ex_ATN}\textbf{A graphical representation of an atom transition
network for the DOPA decarboxylase reaction.} Nodes (open circles)
represent atoms. Atoms can be matched to metabolite structures in
Fig. \ref{fig:ex_met} on their metabolite identifiers, colours and
numbers. Directed edges (arrows) represent atom transitions. All except
one hydrogen atom are omitted to simplify the figure.}
\end{figure}

Metabolic reactions conserve mass and chemical elements. Therefore,
there must exist a mapping from each atom in a reactant metabolite
to a single atom of the same element in a product metabolite. An atom
transition is a single mapping from a substrate to a product atom.
An \textit{atom transition network} contains information about all
atom transitions in a metabolic network. It is a mathematical structure
that enables one to trace the paths of each individual atom through
a metabolic network. An atom transition network can be generated automatically
from a stoichiometric matrix for a metabolic network and atom mappings
for internal reactions. The atom transition network for the DOPA decarboxylase
reaction from the example metabolic network in Fig. \ref{fig:ex_met}
is shown in Fig. \ref{fig:ex_ATN}. Unlike metabolic networks, atom
transition networks are graphs since every atom transition (edge)
connects exactly two atoms (nodes). They are \textit{directed graphs}
since every atom transition has a designated direction that is determined
by the directionality of the parent metabolic reaction, i.e., the
designation of substrates and products. Because atom transition networks
are graphs, they are amenable to analysis with efficient graph algorithms
that are not generally applicable to metabolic networks due to the
presence of hyperedges \cite{klamt_hypergraphs_2009}.

\section{\label{sec:Results}Results}

\subsection{\label{sub:GeneralMethod}Identification of conserved moieties in
the dopamine synthesis pathway}

\begin{figure}[h]
\begin{centering}
\includegraphics{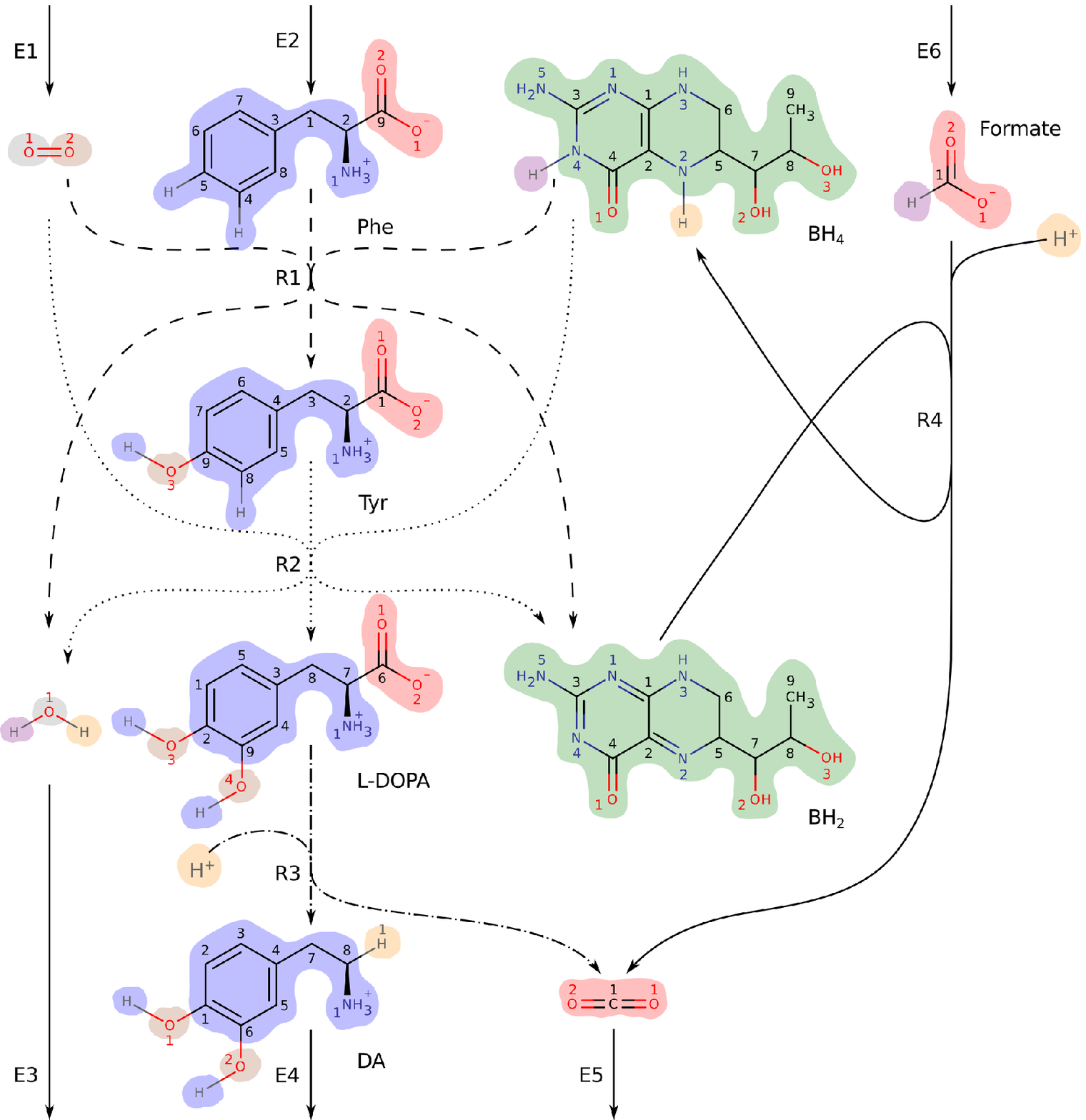}
\par\end{centering}

\protect\caption{\label{fig:DAS_met}\textbf{DAS: a small metabolic network consisting
of reactions in the human dopamine synthesis pathway.} Metabolite
abbreviations are, Phe: L-phenylalanine (KEGG Compound ID: C00079),
Tyr: L-tyrosine (C00082), L-DOPA: levodopa (C00355), DA: dopamine
(C03758), BH\protect\textsubscript{4}: tetrahydrobiopterin (C00272),
BH\protect\textsubscript{2}: dihydrobiopterin (C00268). Internal
reactions are labelled R1-R4. R1 (dashed lines) is the phenylalanine
hydroxylase reaction (KEGG Reaction ID: R01795), R2 (dotted lines)
is the tyrosine hydroxylase reaction (R01815), R3 (dash-dotted lines)
is the DOPA decarboxylase reaction (R02080), and R4 (solid line) is
a composite of the formate dehydrogenase reaction (R00519) and the
dihydropteridine reductase reaction (R01793). Exchange reactions are
labelled E1-E6. The hydrogen ion (H\protect\textsuperscript{\textcolor{blue}{+}})
exchange reaction E7 was omitted to simplify the figure. Atoms are
numbered according to their order in each metabolite's molfile. Atoms
of different elements are numbered separately, in colours matching
their elemental symbol. Atoms belonging to the same conserved moiety
have identically coloured backgrounds.}
\end{figure}

We will demonstrate our method by identifying conserved moieties in
the simple dopamine synthesis network DAS in Fig. \ref{fig:DAS_met}.
This network consists of 11 metabolites, four internal reactions and
seven exchange reactions. The total stoichiometric matrix $S=\left[N,B\right]$
is given in Table \ref{tab:DAS_S}. The internal stoichiometric matrix
$N$ is row rank deficient with ${\rm rank}\left(N\right)=4$. The
dimension of the left null space is therefore $\dim\left({\cal N}\left(N^{T}\right)\right)=7$,
meaning that there are seven linearly independent conservation vectors
for the closed metabolic network. Our analysis of an atom transition
network for DAS will conclude with the computation of seven linearly
independent moiety vectors that span ${\cal N}\left(N^{T}\right)$.
To compute these vectors we require the internal stoichiometric matrix
in Table \ref{tab:DAS_S} and atom mappings for the four internal
reactions. Here, we used algorithmically predicted atom mappings \cite{first_stereochemically_2012}.
These data are required to generate an atom transition network for
DAS (see Section \ref{sub:BuildATN}).
By graph theoretical analysis of this atom transition network we derive
the first of two alternative representations of moiety conservation
relations which we term \textit{moiety graphs}. Nodes in a moiety
graph represent separate instances of a conserved moiety. Each node
is associated with a specific set of atoms in a particular metabolite.
The second representation of moiety conservation relations are the
moiety vectors which can be derived from moiety graphs in a straightforward
manner.  Moiety vectors computed with our method are therefore associated
with specific atoms via moiety graphs.

\begin{table}[!th]
\protect\caption{\label{tab:DAS_S}\textbf{The total stoichiometric matrix $S=\left[N,B\right]$
for DAS.}}

\begin{centering}
\begin{tabular}{l|ccccc|cccccccc|}
\multicolumn{1}{l}{} &  & R1 & R2 & R3 & R4 & E1 & E2 & E3 & E4 & E5 & E6 & E7 & \multicolumn{1}{c}{}\tabularnewline
\cline{2-2} \cline{14-14} 
Phe &  & -1 & 0 & 0 & 0 & 0 & 1 & 0 & 0 & 0 & 0 & 0 & \tabularnewline
Tyr &  & 1 & -1 & 0 & 0 & 0 & 0 & 0 & 0 & 0 & 0 & 0 & \tabularnewline
L-DOPA &  & 0 & 1 & -1 & 0 & 0 & 0 & 0 & 0 & 0 & 0 & 0 & \tabularnewline
DA &  & 0 & 0 & 1 & 0 & 0 & 0 & 0 & -1 & 0 & 0 & 0 & \tabularnewline
CO\textsubscript{2} &  & 0 & 0 & 1 & 1 & 0 & 0 & 0 & 0 & -1 & 0 & 0 & \tabularnewline
Formate &  & 0 & 0 & 0 & -1 & 0 & 0 & 0 & 0 & 0 & 1 & 0 & \tabularnewline
BH\textsubscript{4} &  & -1 & -1 & 0 & 1 & 0 & 0 & 0 & 0 & 0 & 0 & 0 & \tabularnewline
BH\textsubscript{2} &  & 1 & 1 & 0 & -1 & 0 & 0 & 0 & 0 & 0 & 0 & 0 & \tabularnewline
O\textsubscript{2} &  & -1 & -1 & 0 & 0 & 1 & 0 & 0 & 0 & 0 & 0 & 0 & \tabularnewline
H\textsubscript{2}O &  & 1 & 1 & 0 & 0 & 0 & 0 & -1 & 0 & 0 & 0 & 0 & \tabularnewline
H\textsuperscript{+} &  & 0 & 0 & -1 & -1 & 0 & 0 & 0 & 0 & 0 & 0 & 1 & \tabularnewline
\cline{2-2} \cline{14-14} 
\end{tabular}\bigskip{}

\par\end{centering}

Rows and columns are labelled, respectively, with the corresponding
metabolite and reaction identifiers from Fig. \ref{fig:DAS_met}.
The hydrogen ion (H\textsuperscript{+}) exchange reaction E7 was
omitted from Fig. \ref{fig:DAS_met} for simplification. The first
four columns of $S$ correspond to the internal stoichiometric matrix
$N$ and the last seven columns correspond to the exchange stoichiometric
matrix $B$.
\end{table}

\begin{figure}[h]
\begin{centering}
\includegraphics{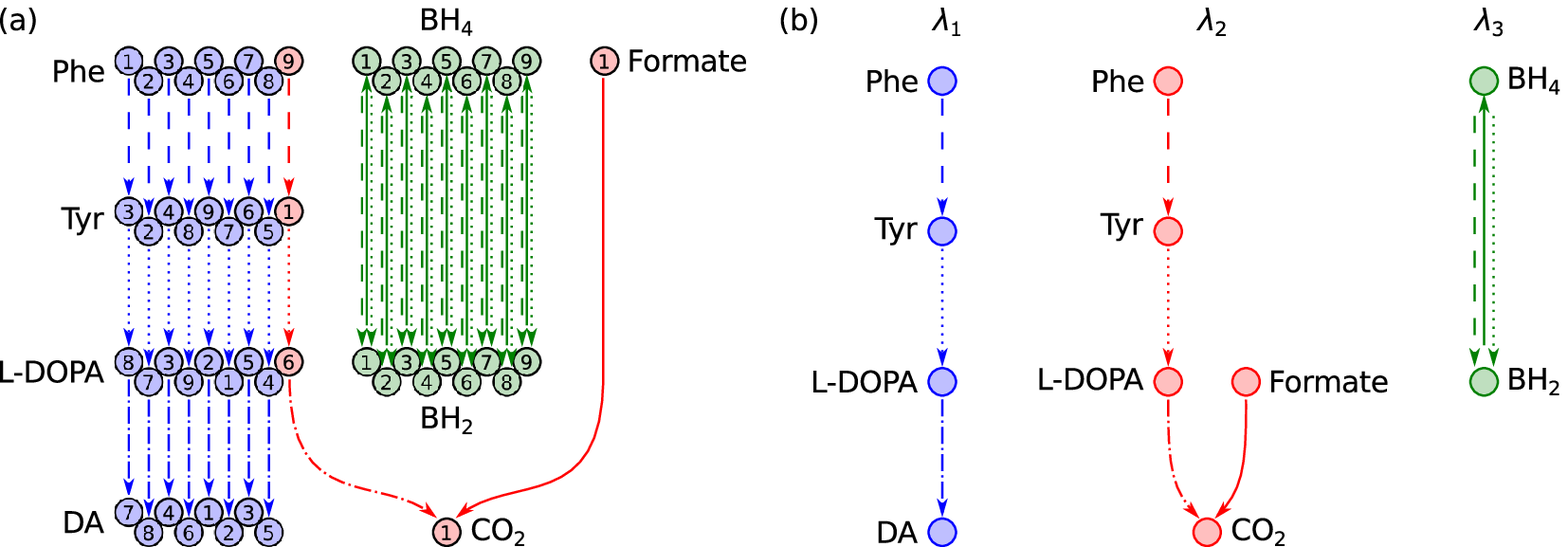}
\par\end{centering}

\protect\caption{\label{fig:DAS_pipeline}\textbf{Identification of conserved carbon
moieties in DAS.} (a) The carbon atom transition network. Numbering
of atoms and line styles of atom transitions refer to Fig. \ref{fig:DAS_met}.
The directed graph consists of 18 components, one for each of the
nine carbon atoms in L-phenylalanine, and one for each of the nine
carbon atoms in tetrahydrobiopterin. The single carbon atom (C1) in
formate is in the same component as C9 in L-phenylalanine, since a
path can be traced between the two atoms when directionalities of
atom transitions are ignored. Isomorphic components have matching
colours. A single instance of a conserved moiety consists of all equivalent
atoms in a set of isomorphic components. (b) Moiety graphs for the
three carbon moieties in DAS. Each graph was obtained by merging a
set of isomorphic components in (a) into a single directed graph.
Each node represents an instance of a conserved moiety. Each edge
represents conservation of a moiety between two metabolites in a particular
reaction. Colours match the background colours of the corresponding
moieties in Fig. \ref{fig:DAS_met}. Analysis of the full atom transition
network for DAS yielded four additional conserved moieties (Fig. \ref{fig:DAS_moieties}).}
\end{figure}

To identify all conserved moieties in DAS we require an atom transition
network for all atoms regardless of element but for demonstration
purposes we will initially focus only on carbon atoms. A carbon atom
transition network for DAS is shown in Fig. \ref{fig:DAS_pipeline}a.
Our working definition of a conserved moiety is a group of atoms that
follow identical paths through a metabolic network. To identify conserved
moieties, we therefore need to trace the paths of individual atoms
and determine which paths are identical. The paths of individual atoms
through the carbon atom transition network for DAS can be traced by
visual inspection of Fig. \ref{fig:DAS_pipeline}a. For example, we
can trace a path from C1 in L-phenylalanine to C7 in dopamine via
C3 in L-tyrosine and C8 in levodopa. This path is made up of atom
transitions in reactions R1, R2, and R3 from Fig. \ref{fig:DAS_met}.
In graph theory terms, these four carbon atoms and the atom transitions
that connect them constitute a\textit{ connected component} \cite{gross_graph_2005}
or, simply, a \textit{component} of the directed graph representing
the carbon atom transition network for DAS. A directed graph is said
to be \textit{connected} if a path exists between any pair of nodes
when edge directions are ignored. A component of a directed graph
is a maximal connected subgraph. In total, the carbon atom transition
network for DAS in Fig. \ref{fig:DAS_pipeline}a consists of 18 components.

The paths of the first eight carbon atoms (C1-C8) in L-phenylalanine
are identical in the sense that they include the same number of atoms
in each metabolite and the same number of atom transitions in each
reaction. In graph theory terms, the components containing C1-C8 in
L-phenylalanine are \textit{isomorphic}. An isomorphism between two
graphs is a \textit{structure preserving} vertex bijection \cite{gross_graph_2005}.
The definition of isomorphism varies for different types of graphs
as they have different structural elements that need to be preserved.
An isomorphism between two simple graphs is a vertex bijection that
preserves the adjacency and nonadjacency of every node, i.e., its
connectivity. An isomorphism between two simple directed graphs must
also preserve edge directions. We define an isomorphism between two
components of an atom transition network as a vertex bijection that
preserves the metabolic identity of every node. The nature of chemical
reactions ensures that all other structural elements are preserved
along with metabolic identities, including the connectivity of atoms
and the number, directions and reaction identities of atom transitions.
The 18 components of the carbon atom transition network for DAS in
Fig. \ref{fig:DAS_pipeline}a can be divided into three sets, where
every pair of components within each set is isomorphic. 

An isomorphism between two components of an atom transition network
is a one-to-one mapping between atoms in the two components. For example,
the isomorphism between the two left-most components in Fig. \ref{fig:DAS_pipeline}a
maps between C1 and C2 in L-phenylalanine, C3 and C2 in L-tyrosine,
C8 and C7 in L-DOPA, and C7 and C8 in dopamine. We say that two atoms
are \textit{equivalent} if an isomorphism maps between them. We note
that our definition of isomorphism only allows mappings between atoms
with the same metabolic identity. Two atoms can therefore only be
equivalent if they are in the same metabolite. Equivalent atoms follow
identical paths through a metabolic network and therefore belong to
the same conserved moiety. In general, \textit{we define a conserved
moiety to be a maximal set of equivalent atoms in an atom transition
network}. To identify conserved moieties, we must therefore determine
isomorphisms between components of an atom transition network to identify
maximal sets of equivalent atoms.

\begin{figure}[h]
\begin{centering}
\includegraphics{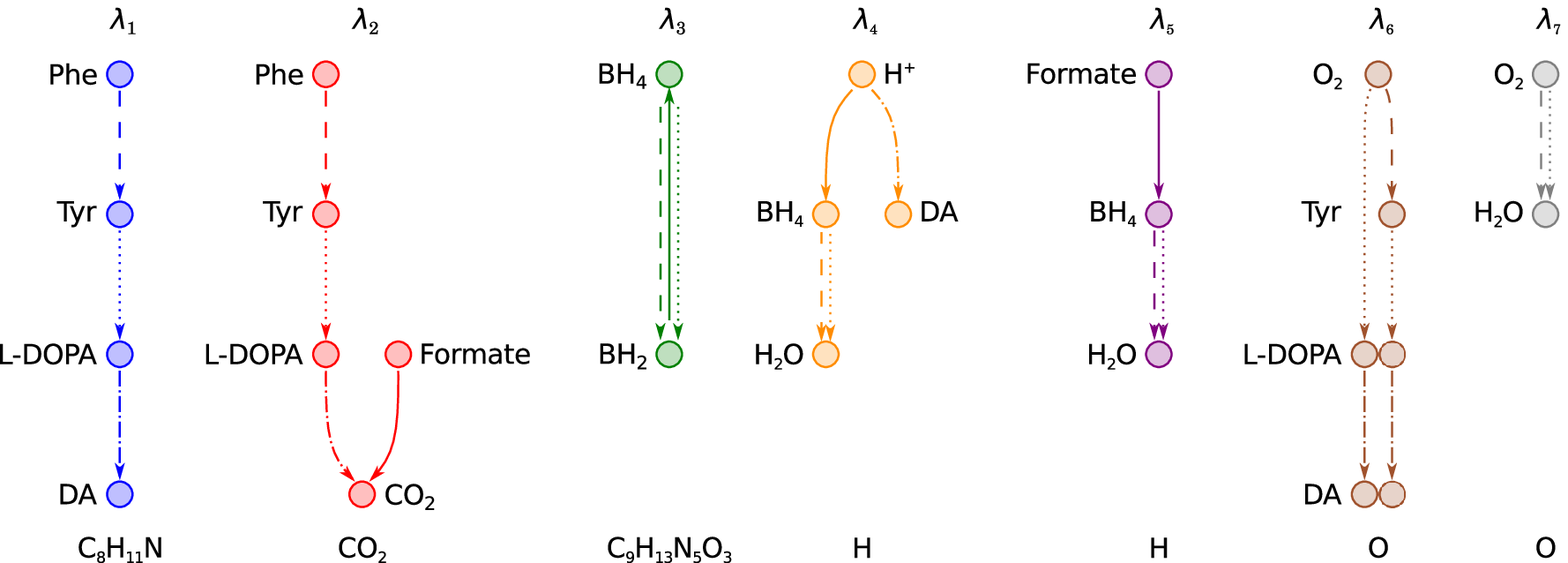}
\par\end{centering}

\protect\caption{\label{fig:DAS_moieties}\textbf{Moiety graphs for all seven conserved
moieties in DAS.} The seven moieties were identified by analysis of
the full atom transition network for DAS in Fig. \ref{fig:DAS_met}.
Colours match the background colours of the corresponding moieties
in Fig. \ref{fig:DAS_met}. The chemical composition of each moiety
is given below its graph. }
\end{figure}

The first eight carbon atoms (C1-C8) in L-phenylalanine are equivalent.
They are therefore part of the same conserved moiety, which we denote
$\lambda_{1}$. The last eight carbon atoms (C2-C9) in L-tyrosine
are likewise part of the same conserved moiety. They make up another
instance of the $\lambda_{1}$ moiety. The $\lambda_{1}$ moiety is
conserved between L-phenylalanine and L-tyrosine in reaction R1, between
L-tyrosine and levodopa in reaction R2, and between levodopa and dopamine
in reaction R3. Each of the four metabolites contains one instance
of the $\lambda_{1}$ moiety. The path of this moiety through DAS
defines its conservation relation. This brings us to our first representation
of moiety conservation relations, which we term moiety graphs. Moiety
graphs are obtained from atom transition networks by merging a set
of isomorphic components into a single graph. Moiety graphs for the
three carbon atom moieties in DAS are shown in Fig. \ref{fig:DAS_pipeline}b.
Four additional moieties were identified by analysis of an atom transition
network for DAS that included all atoms regardless of element. All
seven moiety graphs are shown in Fig. \ref{fig:DAS_moieties}. Atoms
belonging to each node in the moiety graphs are highlighted in Fig.
\ref{fig:DAS_met}.

The second way to represent moiety conservation relations is as moiety
vectors. Above we defined a moiety vector as a conservation vector
$l_{k}$ where element $l_{k,i}$ corresponds to the number of instances
of moiety $k$ in metabolite $i$ of a metabolic network (see Section
\ref{sub:moieties}). We can now make
this definition exact by relating moiety vectors to moiety graphs.
Each instance of a conserved moiety is represented as a node in its
moiety graph. Element $l_{k,i}$ of a moiety vector therefore corresponds
to the number of nodes in moiety graph $\lambda_{k}$ that represent
moieties in metabolite $i$. Moiety vectors are readily derived from
moiety graphs by counting the number of nodes in each metabolite.
Moiety vectors for DAS were derived from the moiety graphs in Fig.
\ref{fig:DAS_moieties}. The seven moiety vectors are given as columns
of the moiety matrix $L\in\mathbb{Z}^{11\times7}$ in Table \ref{tab:DAS_L}.
These seven vectors are linearly independent and therefore span all
seven dimensions of ${\cal N}\left(N^{T}\right)$. The moiety matrix
$L$ is therefore a \textit{moiety basis} for the left null space.

\begin{table}[!th]
\protect\caption{\label{tab:DAS_L}\textbf{Moiety vectors for DAS.}}

\begin{centering}
\begin{tabular}{l|ccccccccc|}
\multicolumn{1}{l}{} &  & $l_{1}$ & $l_{2}$ & $l_{3}$ & $l_{4}$ & $l_{5}$ & $l_{6}$ & $l_{7}$ & \multicolumn{1}{c}{}\tabularnewline
\cline{2-2} \cline{10-10} 
Phe &  & 1 & 1 & 0 & 0 & 0 & 0 & 0 & \tabularnewline
Tyr &  & 1 & 1 & 0 & 0 & 0 & 1 & 0 & \tabularnewline
L-DOPA &  & 1 & 1 & 0 & 0 & 0 & 2 & 0 & \tabularnewline
DA &  & 1 & 0 & 0 & 1 & 0 & 2 & 0 & \tabularnewline
CO\textsubscript{2} &  & 0 & 1 & 0 & 0 & 0 & 0 & 0 & \tabularnewline
Formate &  & 0 & 1 & 0 & 0 & 1 & 0 & 0 & \tabularnewline
BH\textsubscript{4} &  & 0 & 0 & 1 & 1 & 1 & 0 & 0 & \tabularnewline
BH\textsubscript{2} &  & 0 & 0 & 1 & 0 & 0 & 0 & 0 & \tabularnewline
O\textsubscript{2} &  & 0 & 0 & 0 & 0 & 0 & 1 & 1 & \tabularnewline
H\textsubscript{2}O &  & 0 & 0 & 0 & 1 & 0 & 0 & 1 & \tabularnewline
H\textsuperscript{+} &  & 0 & 0 & 0 & 1 & 1 & 0 & 0 & \tabularnewline
\cline{2-2} \cline{10-10} 
\end{tabular}\bigskip{}

\par\end{centering}

The seven moiety vectors, denoted $l_{1}$-$l_{7}$ are written as
columns of the moiety matrix $L$. Note that $L_{3,6}=L_{4,6}=2$
because levodopa ($i=3$) and dopamine ($i=4$) each contain two instances
of the $l_{6}$ moiety (see moiety graph $\lambda_{6}$ in Fig. \ref{fig:DAS_moieties}).
\end{table}

\subsection{\label{sub:Alternatives}Effects of variable atom mappings between
recurring metabolite pairs}

Atom transition networks are generated from atom mappings for internal
reactions of metabolic networks. However, atom mappings for metabolic
reactions are not necessarily unique. Computationally predicted atom
mappings, as used here, are always associated with some uncertainty.
In addition, there can be biochemical variability in atom mappings,
in particular for metabolites containing symmetric atoms. All reactions
of the O\textsubscript{2} molecule, for example, have at least two
biochemically equivalent atom mappings since the two symmetric oxygen
atoms map with equal probability to connected atoms. Different atom
mappings give rise to different atom transition networks that may
contain different moiety conservation relations. For the most part,
we found that varying the set of input atom mappings did not affect
the number of computed moiety conservation relations, only their atomic
structure. An important exception was when atom mappings between the
same pair of metabolites varied between reactions in the same metabolic
network.

\begin{figure}[h]
\begin{centering}
\includegraphics{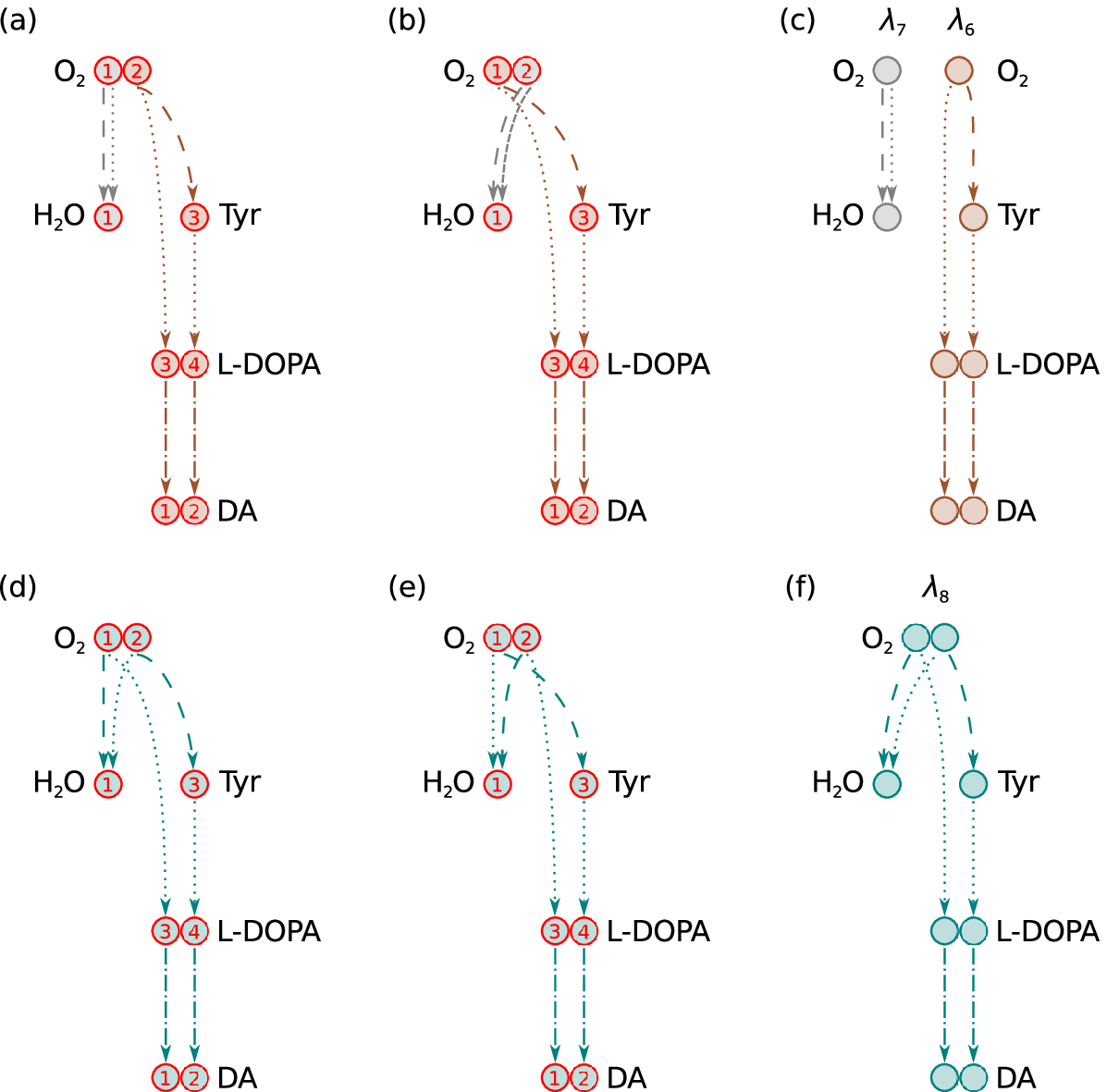}
\par\end{centering}

\protect\caption{\label{fig:DAS_alt}\textbf{Effects of variable atom mappings between
}O\protect\textsubscript{2}\textbf{ and }H\protect\textsubscript{2}O\textbf{
in DAS.} The recurring metabolite pair exchanges an oxygen atom in
two reactions, R1 and R2 in Fig. \ref{fig:DAS_met}. Since the two
oxygen atoms of O\protect\textsubscript{2} are symmetric, there are
four possible combinations of oxygen atom mappings for these two reactions.
Each combination gives rise to a different oxygen transition network.
(a) The first oxygen atom (O1) in O\protect\textsubscript{2} maps
to the single oxygen atom (O1) in H\protect\textsubscript{2}O in
both R1 and R2. (b) O2 in O\protect\textsubscript{2} maps to O1 in
H\protect\textsubscript{2}O in both R1 and R2. (c) Moiety graphs
obtained from the oxygen atom transition networks in (a) and (b).
Two nondecomposable moiety conservation relations were identified
in each atom transition network where the same atom mapped from O\protect\textsubscript{2}
to H\protect\textsubscript{2}O in both R1 and R2. (d) O1 in O\protect\textsubscript{2}
maps to O1 in H\protect\textsubscript{2}O in R1 while O2 in O\protect\textsubscript{2}
maps to O1 in H\protect\textsubscript{2}O in R2. (e) O2 in O\protect\textsubscript{2}
maps to O1 in H\protect\textsubscript{2}O in R1 while O1 in O\protect\textsubscript{2}
maps to O1 in H\protect\textsubscript{2}O in R2. (f) The single moiety
graph obtained from the oxygen atom transition networks in (d) and
(e). Only one composite moiety conservation relation was identified
in each atom transition network where a different atom mapped from
O\protect\textsubscript{2} to H\protect\textsubscript{2}O in R1
than R2.}
\end{figure}

The same pair of metabolites often exchange atoms in multiple reactions
throughout the same metabolic network. Common cofactors such as ATP
and ADP, for example, exchange atoms in hundreds of reactions in large
metabolic networks \cite{thiele_community-driven_2013}. In the dopamine
synthesis network, DAS in Fig. \ref{fig:DAS_met}, O\textsubscript{2}
and H\textsubscript{2}O exchange an oxygen atom in two reactions,
R1 and R2. Since the two oxygen atoms of O\textsubscript{2} are symmetric,
there are four possible combinations of oxygen atom mappings for these
two reactions. Each combination gives rise to a different oxygen transition
network as shown in \ref{fig:DAS_alt}. Two of these oxygen transition
networks, shown in Figures \ref{fig:DAS_alt}a and \ref{fig:DAS_alt}b,
contain two moiety conservation relations each, $\lambda_{6}$ and
$\lambda_{7}$, which are shown in Fig. \ref{fig:DAS_alt}c. The other
two oxygen transition networks, shown in Figures \ref{fig:DAS_alt}d
and \ref{fig:DAS_alt}e, contain only one moiety conservation relation
each, $\lambda_{8}$, which is shown in Fig. \ref{fig:DAS_alt}f.

The DAS atom transition network considered in the previous section
was generated with the oxygen atom mappings in Fig. \ref{fig:DAS_alt}a
and thus contained the two moiety conservation relations $\lambda_{6}$
and $\lambda_{7}$ (see Fig. \ref{fig:DAS_moieties}). An atom transition
network generated with the atom mappings in Fig. \ref{fig:DAS_alt}d
or \ref{fig:DAS_alt}e would contain the single moiety conservation
relation $\lambda_{8}$ instead of these two. What distinguishes the
oxygen transition networks in Figures \ref{fig:DAS_alt}d and \ref{fig:DAS_alt}e
is that the oxygen atom in O\textsubscript{2} that maps to H\textsubscript{2}O
varies between the two reactions R1 and R2. The atom transition network
for DAS therefore contains one less moiety conservation relation if
the atom mapping between this recurring metabolite pair varies between
reactions. The moiety matrix for these alternative atom transition
networks,
\begin{equation}
L=\left[l_{1},l_{2},l_{3},l_{4},l_{5},l_{8}\right],
\end{equation}
only contains six linearly independent columns and is therefore not
a basis for the seven dimensional left null space of $N$.

The vector representation of moiety graph $\lambda_{8}$ is
\begin{equation}
l_{8}^{T}=\left[\begin{array}{ccccccccccc}
0 & 1 & 2 & 2 & 0 & 0 & 0 & 0 & 2 & 1 & 0\end{array}\right].
\end{equation}
We note that $l_{8}=l_{6}+l_{7}$ where 
\begin{equation}
l_{6}^{T}=\left[\begin{array}{ccccccccccc}
0 & 1 & 2 & 2 & 0 & 0 & 0 & 0 & 1 & 0 & 0\end{array}\right],
\end{equation}
\begin{equation}
l_{7}^{T}=\left[\begin{array}{ccccccccccc}
0 & 0 & 0 & 0 & 0 & 0 & 0 & 0 & 1 & 1 & 0\end{array}\right],
\end{equation}
from Table \ref{tab:DAS_L}. The moiety vector $l_{8}$ therefore
represents a composite moiety. It does not meet the definition of
a nondecomposable moiety vector in Eq. \ref{eq:p3}. This example
shows that variable atom mappings between recurring metabolite pairs
may cause multiple nondecomposable moiety conservation relations to
be joined together into a single composite moiety conservation relation.
We formulated an optimisation problem, described in 
Section \ref{sub:Decomposition}, to decompose composite moiety
vectors. Solving this problem for the composite moiety vector $l_{8}$
yields the two nondecomposable components $l_{6}$ and $l_{7}$.

\subsection{\label{sub:Tests}General properties of identified moieties}

We applied our method to identify conserved moieties in three metabolic
networks of increasing size. The networks, listed from smallest to
largest, were the dopamine synthesis network, DAS in Fig. \ref{fig:DAS_met},
the \textit{E. coli} core metabolic network, iCore \cite{orth_reconstruction_2010},
and an atom mapped subset of the generic human metabolic reconstruction,
Recon 2 \cite{thiele_community-driven_2013} which we refer to here
as subRecon. The dimensions of the three networks are given in Table
\ref{tab:Networks}. Further descriptions are provided in 
Section \ref{sub:Networks}. There are seven linearly independent
conservation relations for the closed DAS network, 11 for iCore, and
351 for subRecon.

\begin{table}[!th]
\protect\caption{\label{tab:TheThree}\textbf{Results for the three metabolic networks
treated here.}}

\begin{centering}
\subfloat[\label{tab:Networks}]{

\centering{}%
\begin{tabular}{>{\raggedright}p{5cm}>{\centering}p{2cm}>{\centering}p{2cm}>{\centering}p{2cm}}
\hline 
Network & ~~~DAS & ~~~iCore & ~~~subRecon\tabularnewline
\hline 
Metabolites ($m$) & ~~~11 & ~~~72 & ~~~2,970\tabularnewline
Internal reactions ($u$) & ~~~4 & ~~~74 & ~~~4,261\tabularnewline
${\rm rank}\left(N\right)$ & ~~~4 & ~~~61 & ~~~2,619\tabularnewline
$\dim\left({\cal N}\left(N^{T}\right)\right)$ & ~~~7 & ~~~11 & ~~~351\tabularnewline
\hline 
\end{tabular}}
\par\end{centering}

\begin{centering}
\subfloat[\label{tab:MoietyCounts}]{

\centering{}%
\begin{tabular}{l>{\centering}p{2cm}>{\centering}p{2cm}>{\centering}p{2cm}}
\hline 
\multicolumn{1}{>{\raggedright}p{5cm}}{Initial moieties ($r$)} & ~~~7 & ~~~10 & ~~~345\tabularnewline
${\rm rank}\left(L\right)$ & ~~~7 & ~~~10 & ~~~340\tabularnewline
\multicolumn{1}{l}{Decomposed moieties ($t$)} & ~~~7 & ~~~11 & ~~~353\tabularnewline
${\rm rank}\left(D\right)$ & ~~~7 & ~~~11 & ~~~347\tabularnewline
\hline 
\end{tabular}}
\par\end{centering}

\begin{centering}
\subfloat[\label{tab:MoietyMFA}]{

\centering{}%
\begin{tabular}{>{\raggedright}m{5cm}>{\centering}m{2cm}>{\centering}m{2cm}>{\centering}m{2cm}}
\hline 
Carbon atom isotopomers & ~~~$2.8\times10^{3}$ & ~~~$1.1\times10^{15}$ & ~~~$6.2\times10^{23}$\tabularnewline
Carbon moiety isotopomers & ~~~$2.2\times10^{1}$ & ~~~$1.4\times10^{3}$ & ~~~$4.9\times10^{18}$\tabularnewline
\hline 
\end{tabular}}
\par\end{centering}

\begin{centering}
\subfloat[\label{tab:RunningTimes}]{

\centering{}%
\begin{tabular}{>{\raggedright}p{5cm}>{\centering}p{2cm}>{\centering}p{2cm}>{\centering}p{2cm}}
\hline 
Atoms ($p$) & ~~~170 & ~~~1,697 & ~~~153,298\tabularnewline
Atom transitions ($q$) & ~~~176 & ~~~6,019 & ~~~446,900\tabularnewline
Graph-based method (this work) & ~~~$1.8\times10^{-1}$ & ~~~$5.6\times10^{-1}$ & ~~~$2.8\times10^{2}$\tabularnewline
Vertex enumeration algorithm \cite{avis_pivoting_1992} & ~~~$6.1\times10^{-2}$ & ~~~$9.1\times10^{-1}$ & $>6.0\times10^{5}$\tabularnewline
\hline 
\end{tabular}}
\par\end{centering}

(a) Dimensions of stoichiometric matrices. The number of linearly
independent conservation relations is $\dim\left({\cal N}\left(N^{T}\right)\right)=m-{\rm rank}\left(N\right)$
in a closed network with stoichiometric matrix $N\in\mathbb{Z}^{m\times u}$.
(b) Dimensions of moiety matrices. Initial moiety matrices $L\in\mathbb{N}_{0}^{m\times r}$
were computed directly from predicted atom transition networks. Decomposed
moiety matrices $D\in\mathbb{N}_{0}^{m\times t}$ were derived by
decomposing the columns of $L$ as described in 
Section \ref{sub:Decomposition}. (c) Carbon isotopomers (see
Section \ref{sub:MFA}). Comparison between the number of carbon
atom and carbon moiety isotopomers. (d) Computation times (in seconds)
for the graph-based method presented here, in comparison to the vertex
enumeration algorithm described in \cite{avis_pivoting_1992} (see
Section \ref{sub:Complexity}).
\end{table}

\begin{figure}[h]
\begin{centering}
\includegraphics{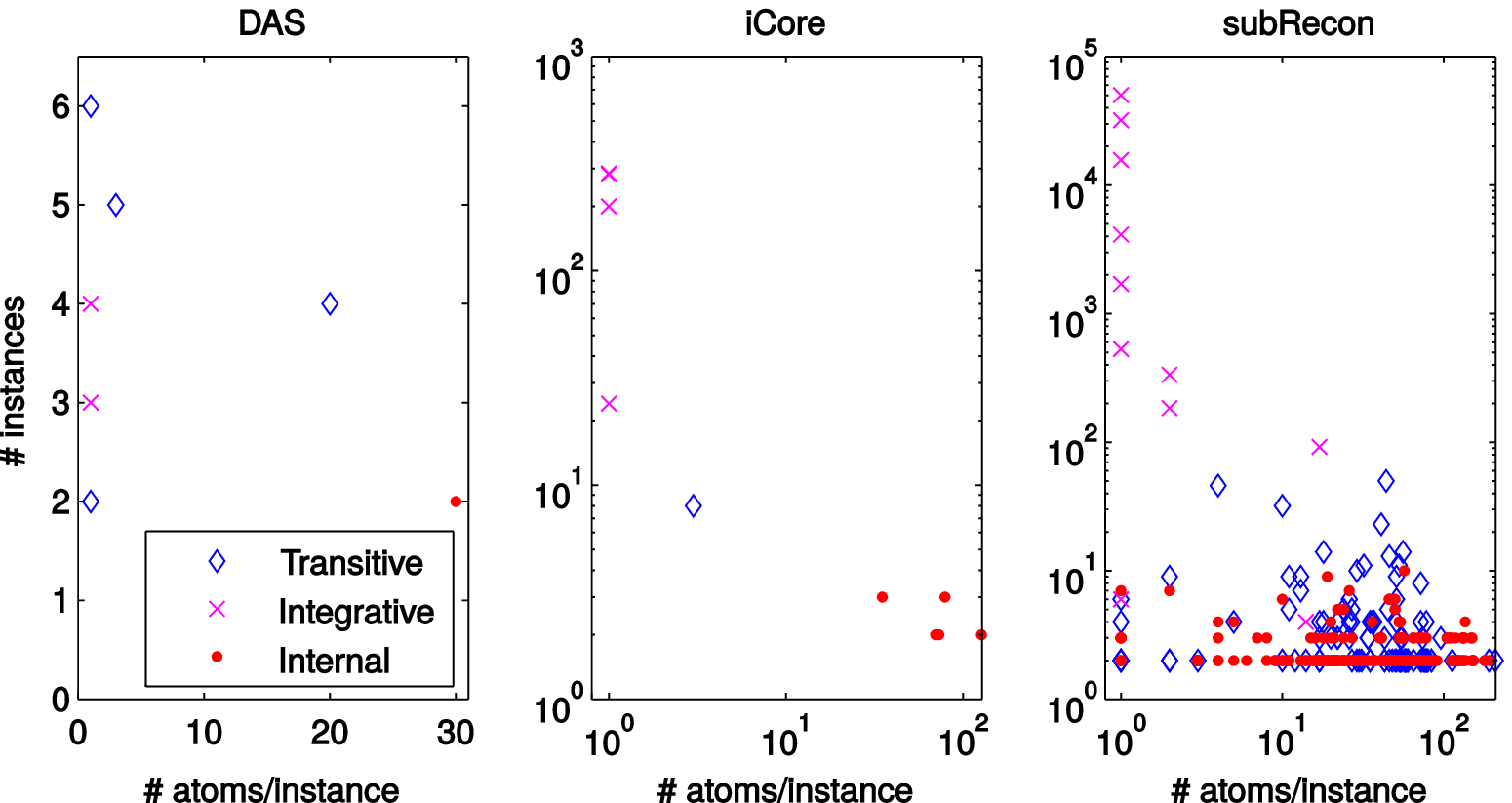}
\par\end{centering}

\protect\caption{\label{fig:Characteristics}\textbf{Characteristics of conserved moieties
identified in the three metabolic networks treated here.} The total
number of instances of a moiety is plotted against the number of atoms
per instance. Classification of moieties as transitive, internal,
or integrative is described in Section \ref{sub:Classification}.}
\end{figure}

Atom transition networks were generated using algorithmically predicted
atom mappings \cite{first_stereochemically_2012} as described in
 Section \ref{sub:BuildATN}. Seven, ten
and 345 moiety conservation relations were identified in the predicted
atom transition network for DAS, iCore and subRecon, respectively
(Table \ref{tab:MoietyCounts}). Characterisation of identified moieties
revealed some trends (Fig. \ref{fig:Characteristics}). We found a
roughly inverse relationship between the frequency of a moiety, defined
as the number of instances, and the size of that moiety, defined as
the number of atoms per instance. We also found a relationship between
moiety size, frequency and classification. Internal moieties tended
to be large and infrequent, occurring in a small number of closely
related secondary metabolites, e.g., the 35 atom AMP moiety found
in the three iCore metabolites AMP, ADP and ATP. Integrative moieties
were usually small and frequent while transitive moieties were intermediate
in both size and frequency. The smallest moieties consisted of single
atoms. These were often highly frequent, occurring in up to 62/72
iCore metabolites and 2,472/2,970 subRecon metabolites. These results
indicate a remarkable interconnectivity between metabolites at the
atomic level. Due to their frequency, single atom moieties accounted
for a large portion of atoms in each metabolic network. Single atom
moieties accounted for nearly half (791/1,697) of all atoms in iCore,
and approximately two thirds (104,268/153,298) of all atoms in subRecon.

Moiety matrices derived from the predicted atom transition networks
for iCore and subRecon did not span the left null spaces of their
respective stoichiometric matrices, indicating that they might contain
composite moiety vectors. Using the method described in 
Section \ref{sub:Decomposition}, we found two composite moiety
vectors in the moiety matrix for iCore, and 10 in the one for subRecon.
Decomposition of these vectors yielded three new nondecomposable moiety
vectors for iCore and 18 for subRecon (Table \ref{tab:MoietyCounts}).
The 11 nondecomposable moiety vectors for iCore were linearly independent.
They therefore comprised a basis for the 11 dimensional left null
space of $N$ for iCore. The 353 nondecomposable moiety vectors for
subRecon, on the other hand, were not linearly independent and only
spanned 347 out of 351 dimensions in the left null space of $S$ for
subRecon. This indicated that there existed conservation relations
for subRecon that were independent of atom conservation.

Schuster and Höfer, citing earlier work by Aris \cite{aris_chemical_1970}
and Corio \cite{corio_theory_1989}, noted the importance of considering
electron conservation in addition to atom conservation \cite{schuster_determining_1991}.
Unfortunately, it is not as straightforward to map electrons as atoms
and no formalism currently exists for electron mappings. As a result,
electron conservation relations cannot be computed with the current
version of our algorithm. We therefore computed electron conservation
relations for subRecon by decomposing the electron vector with the
method described in Section \ref{sub:Decomposition}.
An electron vector for a metabolic network with $m$ metabolites is
a vector $e\in\mathbb{N}^{m}$ where $e_{i}$ is the total number
of electrons in metabolite $i$. Decomposition of $e$ for subRecon
yielded 11 new conservation vectors. When combined, the 11 electron
vectors and the 353 fully decomposed moiety vectors for subRecon (Table
\ref{tab:MoietyCounts}) spanned the left null space of the subRecon
stoichiometric matrix.

\subsection{\label{sub:Gearwheels}The gearwheels of metabolism}

Internal moieties define pools of metabolites with constant total
concentration and dependent individual concentrations. In the small
dopamine synthesis network DAS in Fig. \ref{fig:DAS_met}, the biopterin
moiety ($l{}_{3}$) is classified as internal. This moiety is conserved
between the metabolites BH\textsubscript{2} and BH\textsubscript{4}.
The total concentration of BH\textsubscript{2} and BH\textsubscript{4}
is therefore fixed at a constant value in DAS. If the concentration
of BH\textsubscript{2} increases, the concentration of BH\textsubscript{4}
must decrease by the same amount and vice versa.

The concentration dependency between BH\textsubscript{2} and BH\textsubscript{4}
couples all reactions that interconvert the two metabolites. Assume
that DAS is initially at a steady state when there is a sudden increase
in flux through reactions R1, R2, R3 and associated exchanges such
that the concentrations of all primary metabolites remain constant.
This would lead to net consumption of BH\textsubscript{4} accompanied
by net production of BH\textsubscript{2}. The increased BH\textsubscript{2}/BH\textsubscript{4}
concentration ratio would increase thermodynamic and mass action kinetic
driving forces through R4, while simultaneously decreasing driving
forces through R1 and R2. The system would eventually settle back
to the initial steady state or a new one depending on reaction kinetic
parameters and substrate availability. Conservation of the biopterin
moiety therefore imposes a purely physicochemical form of regulation
on dopamine synthesis that is mediated through mass action kinetics
and thermodynamics. This statement can be generalised to all internal
moieties, as Reich and Sel'kov did in their 1981 monograph on energy
metabolism \cite{reich_energy_1981}.

\begin{figure}[h]
\begin{centering}
\includegraphics{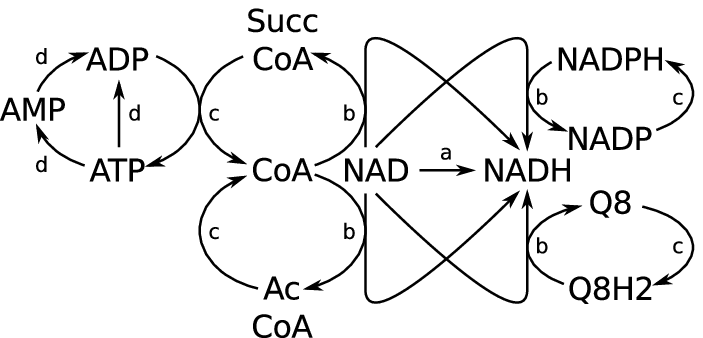}
\par\end{centering}

\protect\caption{\label{fig:iCore_geerwheals}\textbf{Coupling between internal moiety
pools in iCore.} The five pools from Table \ref{tab:TypeC_iCore}
are coupled into a gearwheel-like mechanism. An increase in the NAD/NADH
concentration ratio would affect driving forces in the direction shown.
(a) Any reactions that interconvert NAD and NADH would be driven in
the direction of increased NAD consumption. These include reactions
of glycolysis and the TCA cycle, reactions converting malate and lactate
to pyruvate, and reactions converting pyruvate, ethanol, and acetaldehyde
to acetyl CoA. In short, NAD/NADH coupled reactions would be driven
in the direction of increased acetyl CoA production from available
carbon sources. (b) The increased NAD/NADH concentration ratio would
also affect driving forces through reactions that couple the NAD pool
to other cofactor pools. Altered flux through these reactions would
in turn affect concentration ratios within those pools which are coupled
to their own sets of reactions. (c) An increased NADP/NADPH ratio
would drive flux through the pentose phosphate pathway and conversion
of glutamate to alpha-ketoglutarate. An increased Q8/Q8H2 ratio would
inhibit flux through the electron transport chain. Increased acetyl-CoA/CoA
and succinyl-CoA/CoA ratios would drive acetate production and TCA
cycle reactions, respectively, which are coupled to ATP production
from ADP. (d) An increase in the ATP/ADP ratio resulting from increased
flux through these reactions would drive ATP consuming reactions.
In iCore, ATP consuming reactions are mainly found in gluconeogenesis
so the increased ATP/ADP ratio would counteract the effects of an
increased NAD/NADH ratio to some extent.}
\end{figure}

Reich and Sel'kov's gearwheel analogy \cite{reich_energy_1981} is
appropriate for the five internal moieties we identified in iCore.
These five moieties define five well known cofactor pools (Table \ref{tab:TypeC_iCore}).
Each pool is coupled to a set of reactions that interconvert metabolites
within that pool. The five pools are also coupled to each other through
shared reactions, forming a gearwheel-like mechanism (Fig. \ref{fig:iCore_geerwheals}).
A change in concentration ratios within any pool will affect the driving
forces that turn the wheels. The central wheel in iCore is the NAD
moiety ($l_{6}$). A change in concentration ratios within one pool
will therefore be propagated to other pools via the NAD/NADH concentration
ratio (Fig. \ref{fig:iCore_geerwheals}). This example shows how local
changes in the state of a metabolic network can be propagated throughout
the network via coupled cofactor pools defined by internal moieties.

\begin{table}[!th]
\protect\caption{\label{tab:TypeC_iCore}\textbf{Internal moieties in iCore.}}

\centering{}%
\begin{tabular}{cll}
\hline 
Moiety & Chemical composition & Metabolites\tabularnewline
\hline 
$l_{1}$ & C$_{49}$H$_{74}$O$_{4}$ & Q8, Q8H2\tabularnewline
$l_{2}$ & C$_{21}$H$_{25}$N$_{7}$O$_{17}$P$_{3}$ & NADP, NADPH\tabularnewline
$l_{4}$ & C$_{10}$H$_{12}$N$_{5}$O$_{7}$P & AMP, ADP, ATP\tabularnewline
$l_{6}$ & C$_{21}$H$_{26}$N$_{7}$O$_{14}$P$_{2}$ & NAD, NADH\tabularnewline
$l_{7}$ & C$_{21}$H$_{31}$N$_{7}$O$_{16}$P$_{3}$S & CoA, Acetyl-CoA, Succinyl-CoA\tabularnewline
\hline 
\end{tabular}\bigskip{}
\end{table}

The majority of moieties identified in subRecon were classified as
internal (237/345). Most of these internal moieties were artefacts
of the way the subset of reactions from Recon 2 were selected, i.e.,
based on the availability of atom mapping data (see 
Section \ref{sub:Networks}). Many reactions in subRecon were
disconnected from the rest of the network and therefore could not
carry any flux. To identify reactions capable of carrying flux, we
computed the flux consistent part of subRecon \cite{vlassis_fast_2014},
which consisted of 3,225 reactions and 1,746 metabolites. We identified
118 moiety conservation relations for this part of subRecon, 33 of
which were classified as internal. The metabolite pools defined by
these moieties consisted of between 2 and 9 metabolites and were distributed
across five cell compartments; the cytosol, mitochondria, nucleus,
endoplasmic reticulum, and peroxisomes. Some moieties were compartment
specific while others were distributed amongst metabolites in two
different compartments. As in iCore, the internal moiety pools were
not independent of each other but were coupled by shared reactions.

\subsection{\label{sub:MFA}Application of moiety graphs to stable isotope assisted
metabolic flux analysis}

Atoms in the same instance of a conserved moiety all follow the same
path through a metabolic network. In an atom transition network these
atoms are represented as separate nodes and their atom transitions
as separate edges. A moiety graph encodes the paths of all atoms in
an atom transition network in a reduced number of nodes and edges.
In effect, they are reduced representations of atom transition networks
that can be used in many of the same applications. \foreignlanguage{english}{Atom
transition networks arise most frequently in the context of stable
isotope assisted metabolic flux analysis where they underpin the ability
to model the flow of isotopically labelled atoms through metabolic
networks }\cite{wiechert_13c_2001}\foreignlanguage{english}{. }Stable
isotope assisted metabolic flux analysis (MFA) deals with estimation
of internal reaction fluxes in a metabolic network based on data from
isotope labelling experiments \cite{wiechert_13c_2001}. Internal
fluxes are estimated by fitting a mathematical model to measured exchange
fluxes and isotopomer distributions.

A basic MFA model consists of nonlinear flux balance equations formulated
around isotopomers of metabolites in the metabolic network of interest
\cite{wiechert_bidirectional_1997}. A metabolite with $n$ carbon
atoms has $2^{n}$ carbon atom isotopomers. Therefore, the number
of isotopomer balance equations grows exponentially with the number
of metabolites in the metabolic network. More sophisticated MFA modelling
frameworks have been developed to reduce the complexity of the problem,
notably the cumomer \cite{wiechert_bidirectional_1999} and elementary
metabolite unit (EMU) \cite{antoniewicz_elementary_2007} frameworks.
Cumomer models consist of flux balance equations formulated around
transformed variables called cumomers. They are the same size as isotopomer
models but have a simpler structure that makes them easier to solve.
EMU models have a similar structure as cumomer models but are significantly
smaller. They consist of flux balance equations formulated around
transformed variables known as EMU species. The number of EMU species
for a given metabolic network is much smaller than the number of isotopomers
and cumomers.

MFA models can be derived from moiety graphs instead of atom transition
networks without loss of predictive capacity. We say that a moiety
is labelled if any of its atoms are labelled and define moiety isotopomers
as different labelling states of a metabolite's moieties. The eight
carbon containing metabolites in DAS (Fig. \ref{fig:DAS_met}) have
2,820 possible carbon atom isotopomers. Their 55 carbon atoms can
be grouped into 11 carbon moieties (Fig. \ref{fig:DAS_pipeline}b)
with only 22 possible carbon moiety isotopomers. The reduction in
number of isotopomers is even more pronounced for the two larger metabolic
networks (Table \ref{tab:MoietyMFA}), reaching 12 orders of magnitude
for iCore. It was less for subRecon where a greater proportion of
moieties consist of a single atom (Figure \ref{fig:Characteristics}).
However, it was still substantial. Deriving MFA models from moiety
graphs can therefore reduce the number of model equations by several
orders of magnitude. Isotopomer and cumomer models, in particular,
can be simplified with this approach. The algorithm to generate EMU
species from atom transition networks ensures that atoms in the same
instance of a conserved moiety are always part of the same EMU species.
EMU models derived from moiety graphs will therefore be identical
to those derived from atom transition networks (see supporting file
\nameref{sub:S2_Figure}). Regardless of the MFA modelling framework,
moiety graphs can be used to simplify design of isotope labelling
experiments, by reducing the number of options for labelled substrates.

\subsection{\label{sub:Subnetworks}Application of moiety vectors to decomposition
of metabolic networks}

\begin{figure}[h]
\begin{centering}
\includegraphics{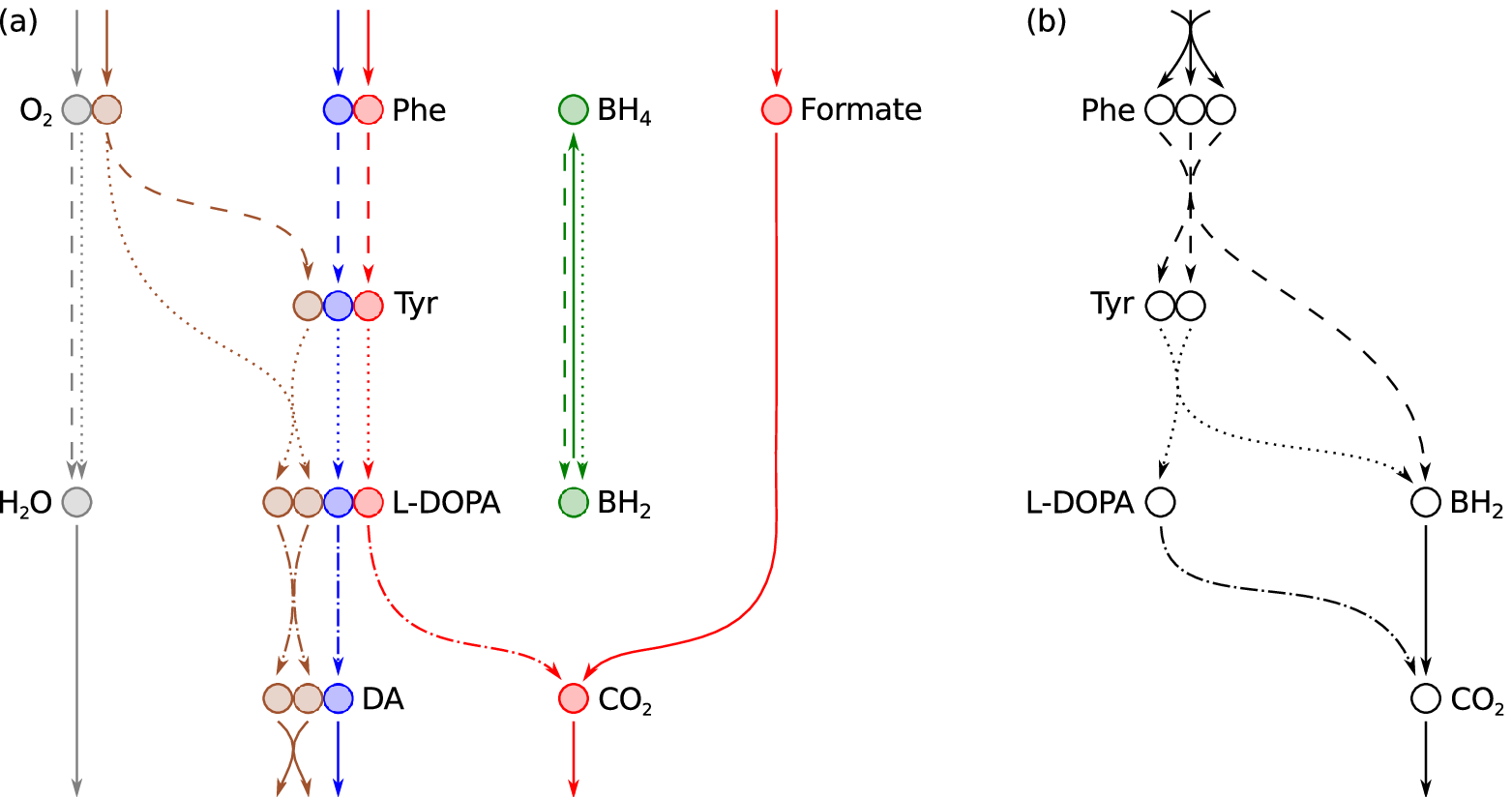}
\par\end{centering}

\protect\caption{\label{fig:DAS_decomposed}\textbf{Moiety subnetworks of DAS.} (a)
Moiety vectors $l_{1}$, $l_{2}$, $l_{3}$, $l_{6}$, and $l_{7}$
(Table \ref{tab:DAS_L}) were used to decompose the stoichiometric
matrix for DAS (Table \ref{tab:DAS_S}) into five subnetworks. Colours
match the corresponding moieties in Fig. \ref{fig:DAS_met} and \ref{fig:DAS_moieties}.
The two hydrogen atom moiety subnetworks ($l_{4}$ and $l_{5}$) were
omitted to simplify the figure. (b) A subnetwork derived from an extreme
ray that did not represent moiety conservation. This subnetwork is
not mass balanced as there is no mass transfer between Phe and BH\protect\textsubscript{\textcolor{blue}{2}},
Tyr and BH\protect\textsubscript{\textcolor{blue}{2}}, or BH\protect\textsubscript{\textcolor{blue}{2}}
and CO\protect\textsubscript{\textcolor{blue}{2}} in the full metabolic
network (Fig. \ref{fig:DAS_met} ).}
\end{figure}

Moiety vectors can be used to decompose a metabolic network into simpler
moiety subnetworks \cite{plasson_decomposition_2008}. An open metabolic
network with total stoichiometric matrix $S$ can be decomposed into
$t$ moiety subnetworks where $t$ is the number of moiety conservation
relations for the corresponding closed network $N$. Each moiety vector
$l_{k}\in{\cal N}\left(N\right)$ defines a stoichiometric matrix
for one moiety subnetwork as 
\begin{equation}
S^{\left(k\right)}=diag\left(l_{k}\right)S.
\end{equation}
Stoichiometric matrices for moiety subnetworks ($S^{\left(k\right)}$)
are generally more sparse than the stoichiometric matrix for the full
metabolic network ($S$). Each moiety subnetwork only includes the
subparts of metabolites and reactions that involve a particular moiety.
Moiety subnetworks of DAS are shown in Fig. \ref{fig:DAS_decomposed}a.
In addition to being more sparse than the full metabolic network (Fig.
\ref{fig:DAS_met}), these subnetworks have simpler topologies. Of
the seven moiety subnetworks of DAS only one ($S^{\left(6\right)}$)
was a hypergraph. All other DAS subnetworks were graphs. Four of 11
iCore subnetworks and 342 of 365 subRecon subnetworks were also graphs.
We note that, although metabolic networks could in theory be decomposed
with other types of conservation vectors, only moiety vectors are
guaranteed to result in mass balanced subnetworks (see Fig. \ref{fig:DAS_decomposed}b).

\subsection{Instantaneous moieties}

The results above were for moieties identified for metabolic network
reconstructions where we assume each reaction is active. These moieties
will only be relevant if all reactions in those reconstructions are
actually active in practice, i.e., carrying nonzero flux. In general,
not all reactions in a metabolic network are active simultaneously,
e.g., oxidative phosphorylation reactions in iCore are only active
in the presence of oxygen. The set of instantaneous conserved moieties,
their conservation relations, and their classification depend on which
reactions are active at any point in time. All steady state flux distributions
$v\in\mathbb{R}^{n}$ are in the right null space ${\cal N}\left(S\right)$
of the total stoichiometric matrix $S$ for a metabolic network \cite{palsson_systems_2015}.
A convex basis for ${\cal N}\left(S\right)$ gives all extreme pathways
of a metabolic network \cite{schilling_theory_2000}. Extreme pathways
are analogous to extreme semipositive conservation relations in the
left null space ${\cal N}\left(S^{T}\right)$ (see Section \ref{sec:Introduction}).
They are a maximal set of conically independent steady state flux
distributions. Any steady state flux distribution can be written as
a conical combination of extreme pathways.

\begin{figure}[h]
\begin{centering}
\includegraphics{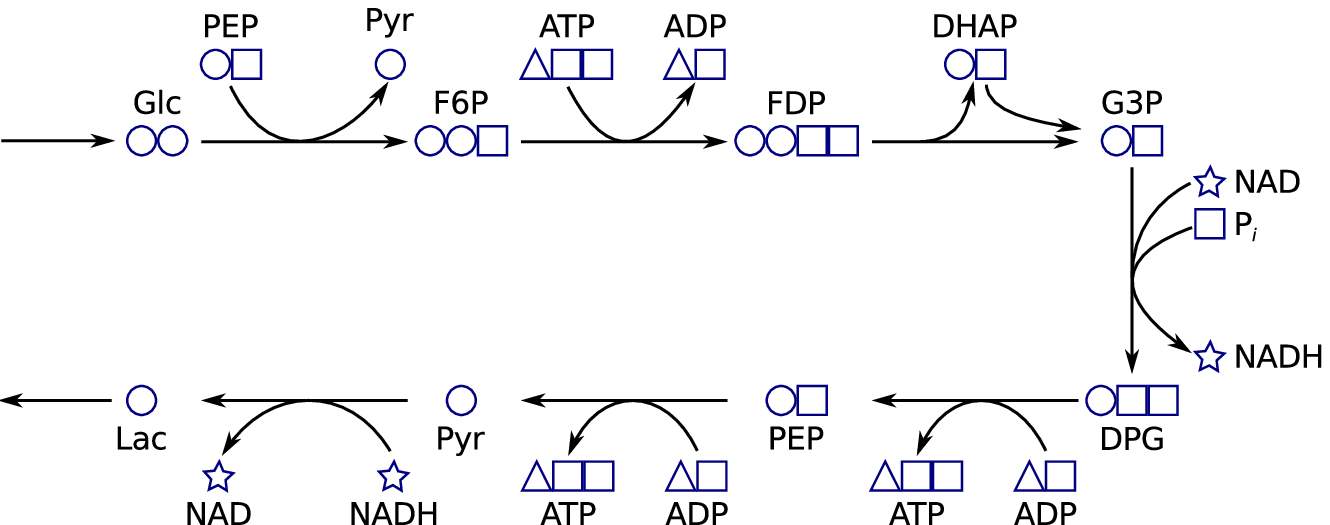}
\par\end{centering}

\protect\caption{\textbf{\label{fig:instantaneous}Instantaneous iCore moieties.} Carbon
and phosphate containing moieties in an extreme pathway of the \textit{E.
coli} core network that corresponds to glycolysis. Four conserved
moieties are distinguished by shape in the figure. The pathway also
conserves one oxygen atom moiety and two hydrogen atom moieties that
were omitted to simplify the figure. Metabolite abbreviations are,
Glc: D-glucose (KEGG Compound ID: C00031), PEP: phosphoenolpyruvate
(C00074), Pyr: pyruvate (C00022), F6P: D-fructose 6-phosphate (C00085),
ATP: adenosine triphosphate (C00002), ADP: adenosine diphosphate (C00008),
FDP: D-fructose 1,6-bisphosphate (C00354), DHAP: dihydroxyacetone
phosphate (C00111), G3P: glyceraldehyde 3-phosphate (C00661), NAD:
nicotinamide adenine dinucleotide (C00003), P\textit{\protect\textsubscript{\textcolor{blue}{i}}}:
orthophosphate (C00009), NADH: reduced nicotinamide adenine dinucleotide
(C00004), DPG: 1,3-bisphospho-D-glycerate (C00236), Lac: D-lactate
(C00256). The glucose moiety (circles) is transitive whereas the other
three moieties are internal, including the phosphate moiety (squares)
which was classified as integrative in the full iCore network. }
\end{figure}

To see how instantaneous conserved moieties vary depending on what
reactions are active we computed the extreme pathways of iCore with
the vertex enumeration algorithm from \cite{avis_pivoting_1992}.
Computation of the extreme pathways of subRecon with the same algorithm
was not tractable. The algorithm returned 1,421 extreme pathways for
iCore. The number of instantaneous moiety conservation relations for
these pathways ranged from 4 to 11 and the total number of moieties
(i.e., instances) ranged from 18 to 520. Figure \ref{fig:instantaneous}
shows an example of instantaneous moieties in an extreme pathway that
corresponds to glycolysis. We found that moieties classified as transitive
or integrative in the entire iCore network, were often classified
as internal in individual extreme pathways. In particular, the inorganic
phosphate moiety (P$_{i}$) was classified as internal in all except
one extreme pathway. The constant metabolite pool defined by the P$_{i}$
moiety varied between pathways, consisting of P$_{i}$, ATP, AMP and
9 to 17 phosphorylated intermediates of glycolysis and the pentose
phosphate pathway. The ammonia moiety (NH\textsubscript{4}\textsuperscript{+})
was also classified as internal in many extreme pathways (266/1,421)
where it defined a constant metabolite pool consisting of NH\textsubscript{4}\textsuperscript{+},
glutamine and glutamate.

\subsection{\label{sub:Complexity}Computational complexity}

The computational complexity of the method presented here is largely
determined by the following two steps: 1) finding connected components
of an atom transition network, and 2) determining isomorphisms between
components. We used an implementation of Tarjan's Algorithm \cite{tarjan_depth_1972}
to find connected components of atom transition networks (see 
Section \ref{sub:CoreMethod}). The worst case time complexity
of this algorithm is $O\left(p+q\right)$ where $p$ is the number
of atoms (nodes) and $q$ is the number of atom transitions (edges)
in the input atom transition network. We apply Tarjan's algorithm
to the simple graph underlying the input atom transition network,
which generally contains significantly fewer edges.

Algorithms to determine isomorphisms between two general graphs are
an active research area. Atom transition networks are specialised
graphs where every node is associated with a metabolite and every
edge is associated with a reaction in the parent metabolic network.
These additional structural elements of atom transition networks make
it possible to determine isomorphisms between their components by
pairwise comparisons (see Section \ref{sub:CoreMethod}).
Since every atom must be connected to at least one other atom, the
number of components is bounded from above by $\nicefrac{p}{2}$.
The number of components in the atom transition networks treated here
was much lower. There were 57 components in the atom transition for
DAS, 391 in the one for iCore, and 16,348 in the one for subRecon.
If no component is isomorphic to any other component, we need to compare
the first component to all other components, the second component
to all others except the first, etc. The maximum number of comparisons
is therefore
\begin{equation}
\left(\frac{p}{2}-1\right)+\left(\frac{p}{2}-2\right)+\cdots+\left(\frac{p}{2}-\frac{p}{2}\right)=\frac{p^{2}}{4}-\sum_{g=1}^{\nicefrac{p}{2}}g=\frac{p^{2}}{4}-\frac{1}{2}\left(\frac{p^{2}}{4}+\frac{p}{2}\right)=\frac{1}{4}\left(\frac{p^{2}}{2}-p\right).
\end{equation}
The overall worst case time complexity of our method is therefore
$\mathcal{O}\left(p^{2}+q\right)$. In practice, however, computation
time scales much better (Table \ref{tab:RunningTimes}). Identification
of conserved moieties in subRecon took under five minutes with our
method. We compared this performance with an implementation of a vertex
enumeration algorithm \cite{avis_pivoting_1992} to compute the extreme
rays of the left null space of a stoichiometric matrix (Table \ref{tab:RunningTimes}).
The two algorithms performed similarly on the two smaller networks
but computation of extreme rays proved intractable for subRecon. The
vertex enumeration algorithm did not complete after running for a
week, at which point we terminated the process.

It may be of interest to know how our method scales with the size
of metabolic networks, instead of the size of atom transition networks.
The number of atoms per metabolite varies greatly but is bounded from
above. So is the number of atom transitions per reaction. The largest
metabolite in the three metabolic networks treated here was the subRecon
metabolite neurotensin (Recon 2 ID C01836), with 241 atoms. The largest
reaction was the subRecon reaction peroxisomal thiolase 2 (Recon 2
ID SCP2x), with 1,791 atom transitions. This is a composite reaction
with large stoichiometric coefficients. Such large reactions are anomalous.
The average number of atom transitions per metabolic reaction was
much lower. The average ($\pm$standard deviation) was 44 ($\pm$16)
for DAS, 81 ($\pm$72) for iCore, and 105 ($\pm$90) for subRecon.
The number of atoms and atom transitions scales approximately linearly
with the number of metabolites and internal reactions, respectively
(Table \ref{tab:RunningTimes}). We can therefore approximate the
worst case time complexity of our method as $\mathcal{O}\left(m^{2}+u\right)$.

\section{Discussion}

Moiety conservation relations are a subset of nonnegative integer
conservation relations for a metabolic network. In principle, the
latter can be computed using only a stoichiometric matrix, but the
computational complexity of existing algorithms \cite{schuster_determining_1991,schuster_what_1995,famili_convex_2003,nikolaev_elucidation_2005,de_martino_identifying_2014}
has prohibited their application to large networks. Computation of
moiety conservation relations requires information about the paths
of atoms through metabolic networks in addition to reaction stoichiometry
(see Section \ref{sub:moieties}). Here,
we incorporated this information in the form of atom transition networks.
Doing so allowed us to formulate the problem of computing moiety conservation
relations as a graph theory problem that is solvable in polynomial
time. We related atom paths to connected components of atom transition
networks and conserved moieties to equivalent nodes of isomorphic
components. We provided a novel definition of isomorphism that is
specific to the structure of atom transition networks. This definition
enabled us to determine isomorphisms and identify conserved moieties
in a fast and reliable way. The relationship between conservation
relations and metabolite substructures has long been known \cite{atkinson_cellular_1977,reich_energy_1981,park_jr._complete_1986}.
A relationship between conservation relations and graph theoretical
properties of atom transition networks has not, to our knowledge,
been demonstrated prior to this work. This is also, to our knowledge,
the first polynomial time method to compute nonnegative integer conservation
relations for metabolic networks.

Our method requires data on reaction stoichiometry and atom mappings
for internal reactions of a metabolic network. Reliable data on reaction
stoichiometry are readily available from high quality, manually curated
metabolic network reconstructions that have been published for hundreds
of organisms over the past couple of decades or so. These reconstructions
are accessible in a standardised format \cite{hucka_systems_2003},
e.g., through the BioModels database \cite{li_biomodels_2010}. Atom
mapping data are increasingly becoming accessible through biochemical
databases but are still largely algorithmically generated \cite{latendresse_accurate_2012,kumar_clca:_2014}.
KEGG \cite{hattori_development_2003,kotera_computational_2004} and
BioPath (Molecular Networks GmbH, Erlangen, Germany) provide manually
curated atom mappings but the data are not freely accessible. No database
currently provides mappings for hydrogen atoms or electrons which
are required to compute all conserved moieties in a metabolic network.
Data formats vary between databases as there is currently no agreed
standard. However, the availability and quality of atom mapping data
are rapidly increasing and we expect these issues will be remedied
in the near future.

We chose to use the DREAM algorithm \cite{first_stereochemically_2012}
to predict atom mappings for this work. Advantages of DREAM include
ease of use, the ability to map hydrogen atoms, and use of the information-rich
rxnfile format. A disadvantage of DREAM is that it uses mixed integer
linear programming (MILP) which has exponential worst case time complexity.
Kumar and Maranas recently published the first polynomial time atom
mapping algorithm, called canonical labelling for clique approximation
(CLCA) \cite{kumar_clca:_2014}. An implementation of this algorithm
has not yet been released but should further speed up the process
of obtaining atom mapping predictions. CLCA predictions for 27,000
reactions are already accessible through the MetRxn database \cite{kumar_clca:_2014}.
These predictions were not yet suitable for this work, however, as
they do not include hydrogen atoms.

Conserved moieties identified with our method depend on input atom
mappings (see Section \ref{sub:Alternatives}).
We showed how variable atom mappings between recurring metabolite
pairs could give rise to a non-maximal set of composite moiety vectors.
Note that composite moieties are a biochemical reality, not just an
artefact of the atom mapping algorithm used. Many metabolite pairs
do have multiple biochemically equivalent atom mappings, each of which
is realised in a living organism. For modelling purposes, however,
it is desirable to identify a maximal number of linearly independent
moiety conservation relations. We therefore formulated an MILP algorithm
for decomposition of composite moiety vectors (
Section \ref{sub:Decomposition}). It would be preferable to construct
the atom transition network with minimal variability in atom mappings
between recurring metabolite pairs to avoid composite moieties altogether.
Doing so would be relatively straightforward if input data included
all alternative atom mappings for reactions. Prediction of alternative
atom mappings with the DREAM algorithm is possible but time consuming,
both due to the longer running times required, and because DREAM outputs
each alternative atom mapping in a separate rxnfile. Some effort is
therefore required to integrate alternative predictions. The CLCA
algorithm outputs alternative atom mapping predictions in a single
file by default and should therefore facilitate identification of
nondecomposable moiety conservation relations. Ultimately, however,
predicted atom mappings need to be manually curated for alternatives.

To span the left null space of Recon 2 we needed to decompose the
electron vector $e\in\mathbb{N}_{0}^{m}$ (
Section \ref{sub:Tests}) with the MILP algorithm described in
 Section \ref{sub:Decomposition}. We note
that this MILP algorithm can also be used to decompose the elemental
matrix for a metabolic network. This is in fact a method for nonnegative
integer factorisation of the elemental matrix, similar to the algorithm
presented in \cite{park_jr._complete_1986}. However, this method
has exponential worst case time complexity. Also, while MILP decomposition
of the elemental matrix returns the chemical composition of moieties
it cannot be used to pinpoint the exact group of atoms in a metabolite
that belong to each moiety. Empirically, we found that MILP decomposition
of the elemental matrices for the three metabolic networks treated
here completed in a reasonable amount of time although it scaled much
worse than analysis of atom transition networks ($3.4\times10^{-1}$
s for DAS, $1.8\times10^{0}$ s for iCore, $4.7\times10^{3}$ s for
subRecon, compare to Table \ref{tab:RunningTimes}). In the absence
of atom mapping data, MILP decomposition of the elemental matrix provides
an alternative way to compute moiety conservation relations for metabolic
networks. For the most part, decomposition of elemental matrices gave
the same set of vectors as analysis of atom transition networks. The
only exception was that decomposition of the elemental matrix for
DAS returned the vector
\begin{equation}
l_{9}^{T}=\left[\begin{array}{ccccccccccc}
0 & 1 & 2 & 0 & 2 & 2 & 0 & 0 & 1 & 0 & 0\end{array}\right],
\end{equation}
in place of the oxygen moiety vector $l_{6}$ in Table \ref{tab:DAS_L}.
We note that $l_{9}=l_{6}+2\left(l_{2}-l_{1}\right)$ does not correspond
to a conserved moiety in DAS.

Here, we highlighted three potential applications of our method; to
identify constant metabolite pools (Section
\ref{sub:Gearwheels}), to model isotope labelling experiments
for metabolic flux analysis (Section \ref{sub:MFA}),
and to decompose metabolic networks (Section
\ref{sub:Subnetworks}). These applications take advantage of
our method's unique ability to identify the exact group of atoms that
correspond to each conserved moiety. As we alluded to in the introduction,
another clear application area is metabolic modelling. A nonnegative
integer basis for the left null space can be used to simplify metabolic
models and to compute a full rank Jacobian which is required for many
computational modelling methods \cite{sauro_conservation_2004,vallabhajosyula_conservation_2006}.
Other applications would include minimisation of intermediate metabolite
concentrations \cite{schuster_minimization_1991}, and computation
of minimal cut sets \cite{klamt_minimal_2004}. We also believe our
method may be of value to theoretical biologists. For example, the
ability to decompose metabolic networks into simpler subnetworks may
facilitate research on physical and mathematical properties that are
otherwise obscured by topological complexity.

\section{\label{sec:Methods}Methods}

\subsection{\label{sub:Networks}Metabolic networks}

We tested our method on three metabolic networks of increasing sizes
(see Table \ref{tab:Networks}), two human and one\textit{ E. coli}
network. The \textit{E. coli} network consisted of core metabolic
pathways including glycolysis, the pentose phosphate shunt, the TCA
cycle, oxidative phosphorylation and fermentation \cite{orth_reconstruction_2010}.
We refer to this network as iCore for abbreviation.

The two human networks were derived from the generic human metabolic
reconstruction Recon 2 \cite{thiele_community-driven_2013}. The smaller
of the two consisted of four internal reactions from the dopamine
synthesis pathway and seven metabolite exchange reactions. We refer
to this network as DAS, and its four internal reactions as R1, R2,
R3, and R4. R1 corresponds to Recon 2 reaction r0399, R2 is a composite
of reactions TYR3MO2 and THBPT4ACAMDASE, R3 corresponds to reaction
3HLYTCL, and R4 is a composite of reactions DHPR and FDH.

The larger human network, which we refer to as subRecon, included
approximately two thirds (4,261/6,691) of internal reactions in Recon
2. This was the largest subset of Recon 2 reactions for which atom
mappings could be predicted at the time of our analysis. For most
of the remaining reactions (2,380/2,430), we were unable to generate
rxnfiles for input to the DREAM server \cite{first_stereochemically_2012}.
For other reactions (50/2,430), the DREAM algorithm timed out or failed
to parse input rxnfiles. Rxnfiles could not be generated for 1,871/2,380
due to lack of information about metabolite structures, and for 509/2,380
reactions because they were not mass or charge balanced.

\subsection{\label{sub:BuildATN}Generation of atom transition networks}

Atom transition networks were generated based on atom mappings for
metabolic reactions. Atom mapping predictions were obtained through
the web interface to the mixed integer linear programming method DREAM
\cite{first_stereochemically_2012}. The objective was set to minimise
the number of bonds broken and formed in each reaction. Reactions
were input to DREAM in rxnfile format (Accelrys, San Diego, CA). Rxnfiles
were written from data on reaction stoichiometry and metabolite structures
in molfile format (Accelrys, San Diego, CA). All hydrogen atoms were
explicitly represented to obtain mappings for hydrogen atoms in addition
to other elements. Care was taken to ensure that hydrogen and charge
balancing of reactions was the same in rxnfiles as in the parent stoichiometric
matrix. This was essential to ensure that computed moiety vectors
were in the left null space of the stoichiometric matrix.

\subsection{\label{sub:CoreMethod}Identification of conserved moieties}

We denote the internal stoichiometric matrix of a metabolic network
by $N\in\mathbb{Z}^{m\times u}$. Conserved moieties in the metabolic
network were identified by analysis of an atom transition network
that was generated as described in \ref{sub:BuildATN}. We denote
the incidence matrix of the input atom transition network by $A\in\left\{ -1,0,1\right\} ^{p\times q}$
where $p$ is the number of atoms and $q$ the number of atom transitions.
The first step in our analysis is to find connected components of
$A$. To this end, we used an implementation of Tarjan's algorithm
\cite{tarjan_depth_1972} (see Section \ref{sub:Implementation}).
We denote the incidence matrix of component $h$ of $A$ by $C^{\left(h\right)}\in\left\{ -1,0,1\right\} ^{x\times y}$.

Each atom in a component belongs to a particular metabolite in the
metabolic network. We define a mapping matrix $M^{\left(h\right)}\in\left\{ 0,1\right\} ^{m\times x}$
that maps atoms to metabolites. It is defined such that $M_{i,g}^{\left(h\right)}=1$
if the atom represented by row $g$ in $C^{\left(h\right)}$ belongs
to the metabolite represented by row $i$ in $N$. Otherwise, $M_{i,g}^{\left(h\right)}=0$.
The component $C^{\left(h\right)}$ represents conservation of a single
atom throughout the metabolic network. We define its atom conservation
vector as
\begin{equation}
a_{h}=M^{\left(h\right)}{\bf 1},\label{eq:atomConservation}
\end{equation}
i.e., it is the column sum of $M^{\left(h\right)}$. Element $a_{h,i}$
is therefore the number of atoms in metabolite $i$ that are in component
$C^{\left(h\right)}$. We define two components $C^{\left(h\right)}$
and $C^{\left(d\right)}$ to be isomorphic if they include the same
number of atoms from each metabolite. It follows that the two components
are isomorphic, with $C^{\left(h\right)}=C^{\left(d\right)}$, if
$a_{h}=a_{d}$. A set of isomorphic components is denoted by $K=\left\{ h,d\mid a_{d}=a_{h}\right\} $.

A moiety graph $\lambda_{k}$ is obtained by merging a set $K$ of
isomorphic components into a single graph. The incidence matrix of
$\lambda_{k}$ is given by 
\begin{equation}
G^{\left(k\right)}=\frac{1}{\left|K\right|}\sum_{h\in K}C^{\left(h\right)}.
\end{equation}
We note that $G^{\left(k\right)}=C^{\left(h\right)}\forall h\in K$
except that the rows of $G^{\left(k\right)}$ represent separate instances
of a conserved moiety instead of atoms. A moiety vector $l_{k}$ is
derived from the incidence matrix $G^{\left(k\right)}$ of a moiety
graph in the same way that the atom conservation vector $a_{h}$ was
derived from the incidence matrix $C^{\left(h\right)}$ of a component
in Eq. \ref{eq:atomConservation}. This is equivalent to setting $l_{k}=a_{h}\forall h\in K$.

\subsection{\label{sub:Classification}Classification of moieties}

We classified moieties according to the schema presented in \cite{famili_convex_2003}.
Briefly, moieties were grouped into three categories termed transitive,
integrative and internal. These categories were referred to as Type
A, Type B, and Type C, respectively, in \cite{famili_convex_2003}.
A moiety with conservation vector $l_{k}$ was classified as internal
if it was conserved in the open metabolic network represented by the
total stoichiometric matrix $S$, i.e., if $S^{T}l_{k}=0$. Metabolites
containing internal moieties were defined as secondary metabolites,
while all other metabolites were defined as primary metabolites. Moieties
that were only found in primary metabolites were classified as transitive
moieties, while those that were found in both primary and secondary
metabolites were classified as integrative moieties.

\subsection{\label{sub:Decomposition}Decomposition of moiety vectors}

Our method for analysing atom transition networks returns $r$ moiety
vectors $\left\{ l_{k}\in\mathbb{N}_{0}^{m}\mid k\in\left[1,r\right]\right\} $
as the columns of the moiety matrix $L\in\mathbb{N}_{0}^{m\times r}$.
As described in Section \ref{sub:Alternatives},
our method may return composite moiety vectors if the input atom transition
network was generated from variable atom mappings between recurring
metabolite pairs. Any composite moiety vector can be written as $l_{k}=x_{k}+y_{k}$,
where $x_{k}$ and $y_{k}$ are nonzero moiety vectors. To decompose
a composite moiety vector $l_{k}$, we solved the mixed integer linear
programming (MILP) problem
\begin{eqnarray}
min &  & {\bf 1}^{T}x_{k},\label{eq:obj}\\
s.t. &  & l_{k}=x_{k}+y_{k},\label{eq:sum}\\
 &  & N^{T}x_{k}=0,\label{eq:c1}\\
 &  & x_{k}\in\mathbb{N}_{0}^{m\times1},\label{eq:c2}\\
 &  & 0<{\bf 1}^{T}x_{k}<{\bf 1}^{T}l_{k}.\label{eq:c3}
\end{eqnarray}
We denote this problem by $P_{k}$. The constraint in Eq. \ref{eq:sum}
defines the solution vectors $x_{k}$ and $y_{k}$ as components of
$l_{k}$. The constraints in Eq. \ref{eq:c1} and \ref{eq:c2} correspond
to Eq. \ref{eq:p1} and \ref{eq:p2} defining nonnegative integer
conservation vectors (see Section \ref{sub:moieties}).
These constraints are implicit for $y_{k}$ due to Eq. \ref{eq:sum}.
The constraint in Eq. \ref{eq:c3}, when combined with Eq. \ref{eq:sum},
ensures that $x_{k}$ and $y_{k}$ are both greater than zero. We
chose to minimise the sum of elements in $x_{k}$ but other objectives
would also work. Problem $P_{k}$ is infeasible for nondecomposable
$l_{k}$. We note that the solution vectors $x_{k}$ and $y_{k}$
might themselves be composite moiety vectors. To fully decompose the
moiety matrix $L$ we must therefore solve $P_{k}$ iteratively until
it is infeasible for all input moiety vectors. This process can be
described with the algorithm,
\begin{enumerate}
\item Input $L\in\mathbb{N}_{0}^{m\times r}$. Initialise $L^{\prime}=L$
and $D=\left[\quad\right]$, where $\left[\quad\right]$ denotes an
empty matrix..
\item Set $r^{\prime}=\dim\left(L_{1,:}^{\prime}\right)$ and $L^{\prime\prime}=\left[\quad\right]$,
where $L_{1,:}^{\prime}$ denotes the first row of $L^{\prime}$.\\
\hspace*{1em}If $r^{\prime}\geq1$, then go to Step 3,\\
\hspace*{1em}else, go to Step 5.
\item For $k=1:r^{\prime}$, \\
\hspace*{1em}denote $l_{k}=L_{:,k}$,\\
\hspace*{1em}solve $P_{k}$.\\
\hspace*{1em}If $P_{k}$ is infeasible, set $D=\left[D,l_{k}\right]$,\\
\hspace*{1em}else, denote the solution of $P_{k}$ by $x_{k}$ and
$y_{k}$ and set $L^{\prime\prime}=\left[L^{\prime\prime},x_{k},y_{k}\right]$.\\
\hspace*{1em}Go to Step 4.
\item Set $L^{\prime}=L^{\prime\prime}$ and go back to Step 2.
\item Output the fully decomposed moiety matrix $D\in\mathbb{N}_{0}^{m\times t}$.
\end{enumerate}
The same algorithm can be used for nonnegative integer matrix factorisation
of an elemental matrix and electron vector for a metabolic network.

\subsection{\label{sub:Implementation}Implementation}

We implemented the method presented here as an algorithmic pipeline
in MATLAB (MathWorks, Natick, MA). This implementation is freely available
as part of the COBRA toolbox \cite{schellenberger_quantitative_2011}
at \url{https://github.com/opencobra/cobratoolbox} (directory topology/conservedMoieties).
Required inputs are an atom transition network and a stoichiometric
matrix for a metabolic network. The method outputs moiety conservation
relations both as moiety graphs and moiety vectors. All graphs are
represented as incidence matrices. Support functions to generate atom
transition networks (Section \ref{sub:BuildATN}), classify moieties
(Section \ref{sub:Classification}) and decompose moiety vectors
(Section \ref{sub:Decomposition}) are included with the core
code. A tutorial on identification of conserved moieties in the dopamine
synthesis network DAS is available at \url{https://github.com/opencobra/cobratoolbox}
(directory topology/conservedMoieties/example), along with necessary
data and MATLAB scripts that run through the example.

To compute the connected components of atom transition networks we
used and implementation of Tarjan's algorithm available as part of
the Bioinformatics Toolbox for MATLAB (MathWorks, Natick, MA). This
toolbox is not included with a standard installation of MATLAB. Users
who do not have the Bioinformatics Toolbox can still run the pipeline
with a free alternative to Tarjan's algorithm to compute components
of atom transition networks. If the Bioinformatics Toolbox is not
installed in the MATLAB path, the pipeline calls a k-Nearest Neighbour
algorithm in the MATLAB Network Routines toolbox by Bounova and Weck
\cite{bounova_overview_2012}. This toolbox is freely available with
the COBRA toolbox. The k-Nearest Neighbour algorithm is considerably
slower than Tarjan's algorithm.

All code in the COBRA toolbox is distributed under a GNU General Public
Licence and we encourage implementations of our method for other platforms
than MATLAB. We have taken care to document and comment our code to
facilitate such efforts.

\section*{Supporting Information}

\subsection*{\label{sub:S1_Appendix}S1 Appendix}

\textbf{Mathematical definitions.} Formal definitions of linear algebra
and graph theory terms used or introduced in this work.

\subsection*{\label{sub:S2_Figure}S2 Figure}

\textbf{Conserved moieties and elementary metabolite units.} Application
of the algorithm presented in \cite{antoniewicz_elementary_2007}
to generate an elementary metabolite unit (EMU) reaction network from
a moiety graph.

\section*{Acknowledgments}

This work was supported by the U.S. Department of Energy, Offices
of Advanced Scientific Computing Research and the Biological and Environmental
Research as part of the Scientific Discovery Through Advanced Computing
program, grant \#DE-SC0010429 and by the Luxembourg National Research
Fund (FNR) through the National Centre of Excellence in Research (NCER)
on Parkinson's Disease.

We thank Vuong Phan, Hoai Minh Le, Matthew DeJongh, Averina Nicolae,
Diana El Assal and Fatima Liliana Monteiro for their helpful comments
on the manuscript.

\end{document}